\newcommand\abs[1]{\left|#1\right|}
\newcommand*{\ditto}{\raisebox{-0.5ex}{\ttfamily"}}

\newcommand*\diff{\mathop{}\!\mathrm{d}}
\documentclass[letterpaper]{aastex}
\usepackage{amsbsy}
\usepackage{amsfonts}
\usepackage{amssymb}
\usepackage{lscape}
\usepackage{graphicx}
\usepackage{graphics}
\usepackage{bm}

\begin{document}

\title{A Simple Method for Modeling Collision Processes in Plasmas with a Kappa Energy Distribution}

\author{M. Hahn\altaffilmark{1} and D. W. Savin\altaffilmark{1}}

\altaffiltext{1}{Columbia Astrophysics Laboratory, Columbia University, 550 West 120th Street, New York, NY 10027 USA}

\date{\today}
\begin{abstract}

	We demonstrate that a nonthermal distribution of particles described by a kappa distribution can be accurately approximated by a weighted sum of Maxwell-Boltzmann distributions. We apply this method to modeling collision processes in kappa-distribution plasmas, with a particular focus on atomic processes important for solar physics. The relevant collision process rate coefficients are generated by summing appropriately weighted Maxwellian rate coefficients. This method reproduces the rate coefficients for a kappa distribution to an estimated accuracy of better than 3\%. This is equal to or better than the accuracy of rate coefficients generated using ``reverse engineering'' methods, which attempt to extract the needed cross sections from the published Maxwellian rate coefficient data and then reconvolve the extracted cross sections with the desired kappa distribution. 
Our approach of summing Maxwellian rate coefficients is easy to implement using existing spectral analysis software. Moreover, the weights in the sum of the Maxwell-Boltzmann distribution rate coefficients can be found for any value of the parameter $\kappa$, thereby enabling one to model plasmas with a time-varying $\kappa$. Tabulated Maxwellian fitting parameters are given for specific values of $\kappa$ from 1.7 to 100. We also provide polynomial fits to these parameters over this entire range. Several applications of our technique are presented, including the plasma equilibrium charge state distribution (CSD), predicting line ratios, modeling the influence of electron impact multiple ionization on the equilibrium CSD of kappa-distribution plasmas, and calculating the time-varying CSD of plasmas during a solar flare. 
			
\end{abstract}

	
\maketitle
	
\section{Introduction}\label{sec:intro}

	Evidence for non-Maxwellian particle distributions has been found in the Earth's magnetosphere \citep[e.g.,][]{Vasyliunas:JGR:1968}, the solar wind \citep[e.g.,][]{Feldman:JGR:1975}, the solar corona \citep[e.g.,][]{Cranmer:ApJ:2014}, and solar flares \citep[e.g,][]{Seely:ApJ:1987,Kasparova:AA:2009,Oka:ApJ:2013}. These nonthermal distributions are commonly characterized in terms of kappa distributions, which resemble a Maxwell-Boltzmann distribution at low energy but fall off as a power law at high energies. 
Kappa distributions are predicted by some statistical mechanical theories to be a natural consequence of systems in which there is on-going heating, such as due to reconnection, shocks, or wave-particle interactions \citep[][and references therein]{Pierrard:SolPhys:2010, Dudik:ApJ:2015}. Thus, kappa distributions are likely to exist in a wide variety of astrophysical systems, even beyond the solar and space plasmas that have been the focus of existing work.
	
	The detection of nonthermal electron energy distributions (EEDs) via spectroscopy would be a powerful diagnostic. In solar physics, it has been argued that the presence of high energy electrons supports nanoflare theories of coronal heating \citep{Testa:Sci:2014}. Measurements that characterize the EED could indicate where such nanoflare heating is occuring. Such measurements may also provide insight into the processes within reconnection that lead to particle acceleration \citep{Benz:ARAA:2010}. \citet{Dudik:ApJ:2015} provide a review of recent results concerning kappa distributions in the solar corona. 
	
	In order to detect the presence of kappa EEDs, or other non-thermal distributions, appropriate atomic data are needed to model spectra and analyze observations. For example, for an optically thin spectral line emitted by a transition from level $s$ to level $f$ of charge state $q$ for element $\mathrm{X}$, the line intensity is given by \citep{Phillips:Book}
\begin{equation}
I_{sf} = \frac{1}{4\pi} \int G_{sf} n_{\mathrm{e}}^{2} \diff{h}, 
\label{eq:intensity}
\end{equation}
where $n_{\mathrm{e}}$ is the electron density and the line element $\diff{h}$ lies along the line of sight. The contribution function $G_{sf}$ describes all of the atomic parameters for the transition and is defined as
\begin{equation}
G_{sf} = \frac{n_{s}(\mathrm{X}^{q+})}{n(\mathrm{X}^{q+})}\frac{n(\mathrm{X}^{q+})}{n(\mathrm{X})}\frac{n(\mathrm{X})}{n(\mathrm{H})} \frac{n(\mathrm{H})}{n_{\mathrm{e}}}\frac{A_{sf}}{n_{\mathrm{e}}}.
\label{eq:Gdef}
\end{equation}
Here, $n_{s}(\mathrm X^{q+})/n(\mathrm{X}^{q+})$ is the relative population of the upper level $s$ for charge state $\mathrm{X}^{q+}$. In collisionally excited plasmas, this level population is determined by the balance among collisional excitation, collisional de-excitation, radiative decays, cascades from the decays of higher energy levels, and recombination and ionization into and out of the various energy levels. The next term, $n(\mathrm{X}^{q+})/n(\mathrm X)$, is the relative abundance for charge state $q$ of element $\mathrm{X}$. This describes the charge state distribution (CSD) of the plasma, which for a collisionally ionized plasma is determined by the balance among electron impact ionization (EII) and electron-ion recombination. The other terms give the elemental abundance of $\mathrm{X}$ relative to hydrogen, $n(\mathrm{X})/n(\mathrm{H})$, and the radiative transition transition rate $A_{sf}$. 

	A problem for interpreting spectra from plasmas with a kappa distribution of electrons is that the necessary atomic data have usually only been reported for Maxwellian distributions. The level population and the CSD both depend on the EED, as well as on the electron density. For a Maxwellian plasma, the influence of the distribution is characterized by the temperature $T$. However, for a kappa-distribution plasma, the level population and CSD also depend on the parameter $\kappa$, which characterizes the degree of non-thermality of the distribution. 

	There are several possible approaches available for obtaining the data needed to model collision processes in kappa-distribution plasmas. The most commonly used method has been to ``reverse-engineer'' the data that have been tabulated for Maxwell-Boltzmann EEDs so as to extract the required cross sections and then reconvolve these extracted cross sections with the desired kappa EED \citep[e.g.,][]{Dzifcakova:SolPhys:1992, Wannawichian:ApJS:2003, Dzifcakova:ApJS:2015}. An alternative method is to approximate the kappa distribution itself as a sum of Maxwellians. With this latter approach, the needed atomic collision rate coefficients can be represented as a simple weighted sum of the Maxwellian rate coefficients. This approach has been used, for example, by \citet{Ko:GRL:1996} to model the CSD in the solar wind. However, summing Maxwellians has been relatively neglected in recent work. 
	
	Here, we demonstrate that, in a general way, kappa distributions can be approximated to a very high accuracy by a weighted sum of Maxwellians. We apply this general ``Maxwellian decomposition'' method to modeling collision processes in kappa-distribution plasmas. With this method, atomic process rate coefficients, generated by summing the appropriately weighted Maxwell-Boltzmann rate coefficients, can reproduce the rate coefficients for a kappa distribution to a level equal to or better than the accuracy obtained by using a reverse-engineering method. Moreover, the Maxwellian decomposition method is highly adaptable and can be readily implemented using standard plasma modeling codes. 
	
	The rest of this paper is organized as follows: In Section~\ref{sec:kappa} we introduce the kappa distribution. Section~\ref{sec:atomic} reviews the reverse-engineering approach to generating atomic data and then presents our methods and fitting parameters for representing kappa distributions as a sum of Maxwellians. In Section~\ref{sec:compare} we compare our results for rate coefficients, CSDs, and predicted line intensities to those obtained using the reverse engineering approach. We then extend our analysis to several new applications, including ascertaining the influence of electron impact multiple ionization on the CSD for a plasma with a kappa distribution, in Section~\ref{sec:eimi}, and describing the time-dependent evolution of a plasma following a change in $\kappa$, in Section~\ref{sec:varykap}. Section~\ref{sec:sum} concludes.

\section{Kappa Distributions}\label{sec:kappa}

	An energy distribution quantifies the fraction of particles having an energy between $E$ and $E+\diff{E}$. The isotropic Maxwellian distribution is given by
\begin{equation}
f(E)=\frac{2}{\sqrt{\pi}}\left(\frac{1}{k_{\mathrm{B}}T}\right)^{3/2}\sqrt{E}\exp\left(\frac{-E}{k_{\mathrm{B}}T}\right),
\label{eq:max}
\end{equation}
where $k_{\mathrm{B}}$ is the Boltzmann constant and $T$ is the temperature. 

	The isotropic kappa energy distribution is given by
\begin{equation}
f_{\kappa}(E) = A_{\kappa} \frac{2}{\sqrt{\pi}} \left(\frac{1}{k_{\mathrm{B}}T_{\kappa}}\right)^{3/2}\sqrt{E}
\left[1 + \frac{E}{(\kappa-3/2)k_{\mathrm{B}}T_{\kappa}}\right]^{-(\kappa+1)}, 
\label{eq:kappafunc}
\end{equation}
with 
\begin{equation}\
A_{\kappa} = \frac{\Gamma(\kappa+1)}{\Gamma(\kappa-1/2)(\kappa-3/2)^{3/2}}.
\label{eq:ak}
\end{equation}
Here $\Gamma$ is the Gamma function. The parameter $\kappa$ ranges from 3/2 to $\infty$ with $\kappa=\infty$ being a Maxwellian distribution and smaller values of $\kappa$ corresponding to an increasingly non-thermal distribution. The kappa temperature $T_{\kappa}$ is defined so that the average energy of the particles is $E_{\mathrm{avg}} = (3/2)k_{\mathrm{B}}T_{\kappa}$, in analogy with the usual Maxwellian temperature. The core of the kappa distribution can be approximated by a Maxwellian of temperature $T = T_{\kappa}(\kappa-3/2)/\kappa$. In order for the magnitude of this Maxwellian function to match that of the kappa function at low energies, the Maxwellian must be scaled by a multiplicative factor of \citep{Oka:ApJ:2013, Dzifcakova:ApJS:2013}
\begin{equation}
C=2.718\frac{\Gamma(\kappa+1)}{\Gamma(\kappa-1/2)} \kappa^{-3/2} \left(1+\frac{1}{\kappa}\right)^{-(\kappa+1)}. 
\label{eq:mkscale}
\end{equation}
This scaling factor can be thought of as representing the ratio of the number of thermal particles relative to the total number of particles. That is, it quantifies the fraction of particles at low energies where the distribution function more closely resembles a Maxwellian. Alternatively, $R_{N}=1-C$ is the fraction of non-thermal particles, which is largest for small $\kappa$ and decreases to zero for large $\kappa$ \citep{Oka:ApJ:2013}. For a small value of $\kappa=1.7$ there is a large fraction of nonthermal particles with $R_{N}=0.4$, while a moderate $\kappa=8$ corresponds to $R_{N}=0.1$, and for a large value such as $\kappa=100$ the fraction of nonthermal particles is only $R_{N}=0.01$. 

\section{Atomic Data for Kappa Distributions}\label{sec:atomic}

	In order to model the spectra from an electron-ionized plasma with a kappa EED it is necessary to determine the rate coefficients for the various atomic processes discussed above, in Section~\ref{sec:intro}. These rate coefficients give the number of reactions per unit volume per unit time. As an illustration, the fractional ion abundance of charge state $q$ is $y_{q} \equiv n(\mathrm{X}^{q+})/n(\mathrm X)$. As a function of time, this is described by
\begin{equation}
\frac{\mathrm{d}y_{q}}{\mathrm{d}t}=n_{\mathrm{e}}
\left[\alpha^{p}_{\mathrm{I},q-p}y_{q-p}+\ldots+\alpha^{1}_{\mathrm{I},q-1}y_{q-1}-\left(\alpha^{1}_{\mathrm{I},q}+\ldots+\alpha^{k}_{\mathrm{I},q}+\alpha_{\mathrm{R},q}\right)y_{q}+\alpha_{\mathrm{R},q+1}y_{q+1}\right],
\label{eq:dydt}
\end{equation} 
where $\alpha^{p}_{\mathrm{I},q}$ is the rate coefficient for $p$-times ionization from charge state $q$ to $q+p$ and $\alpha_{\mathrm{R},q}$ is the recombination rate coefficient from $q$ to $q-1$. The terms on the right of the equals sign represent, from left to right, ionization from lower charge states into $q$, ionization and recombination out of $q$ to other charge states, and recombination from $q+1$ into $q$. For most astrophysical atomic plasmas, the density is low so that three-body recombination is extremely unlikely. We also ignore charge exchange, which is only important at low temperatures of $\sim 10^{4}$~K. 

	All of the needed ionization and recombination rate coefficients depend on the EED. Here and throughout we assume the distribution function to be isotropic. For any process having a cross section as a function of speed $\sigma(v)$ or energy $\sigma(E)$ the rate coefficient $\alpha$ is given by \citep[e.g.,][]{Muller:IJMS:1999}
\begin{equation}
\alpha=\int \sigma(v)v f(v) \diff{v} = \int \sigma(E)\sqrt{\frac{2E}{\mu}}f(E)\diff{E},
\label{eq:ratecoeffdef}
\end{equation}
where $\mu$ is the reduced mass for the collision system. We are considering collisions of electrons with atoms and atomic ions so that $\mu \approx m_{\mathrm{e}}$, although our general results also apply to other types of particle collisions. In the case of a Maxwellian distribution, the rate coefficients are generally a function of $T$. For kappa distributions, the rate coefficients are also a function of $\kappa$. 

	The problem for modeling plasmas with kappa EEDs is that atomic data are usually reported as Maxwellian plasma rate coefficients. If the cross sections themselves were given, it would be straightforward to perform the integral of Equation~(\ref{eq:ratecoeffdef}) to obtain the kappa-distribution rate coefficient. This would, in principal, be the most accurate way to obtain the needed rate coefficients. For some processes, such as electron impact ionization, the cross sections themselves are often available. However, for other processes, especially resonant processes, such as dielectronic recombination, it would be extremely cumbersome to tabulate the cross section in a format that could be incorporated into plasma models. Integration over a Maxwellian distribution smoothes out the complex structure in the cross-section data and allows for a simple parameterization. For this reason, it is Maxwellian atomic data that are usually tabulated, while keeping databases of cross sections is, for now, considered impractical. 

\subsection{Reverse-Engineering Approach}\label{subsec:datarev}

	One method for obtaining rate coefficients for a nonthermal distribution is to extract the appropriate rate coefficients from the Maxwellian-integrated atomic data \citep{Owocki:ApJ:1983, Dzifcakova:SolPhys:1992, Porquet:AA:2001, Wannawichian:ApJS:2003, Hansen:PRE:2004}. The idea is to take the fitting formulae used to tabulate the rate coefficients for Maxwellian distributions and extract from them an approximate cross section, which can then be integrated over a kappa distribution. 
	
	To illustrate this reverse-engineering approach, we will discuss the method for dielectronic recombination (DR). DR is a process in which a free electron approaches an ion, excites a bound electron within the ion, and is simultaneously captured \citep{Muller:Book:2008}. The resulting doubly excited state may relax by autoionization or by emitting a photon. Recombination occurs when the excited state relaxes radiatively to below the ionization threshold of the recombined system. DR occurs at collision energies $E = \Delta E - E_{\mathrm{b}}$, where $\Delta E$ is the electronic core excitation energy of the recombined ion and $E_{\mathrm{b}} \approx 13.6 q^2/n^2$~eV is the bound-state energy of the captured electron into a Rydberg level with principal quantum number $n$ of the ion with initial charge $q$. Because both $\Delta E$ and $E_{\mathrm{b}}$ are quantized, DR is a resonant process. For each electronic state there are an infinite series of resonances corresponding to all possible Rydberg levels for the captured electron. Detailed calculations and measurements of DR have been performed that resolve this complex resonant structure \citep[see e.g.,][and references therein]{Schippers:IRAMP:2010}. 
	
	The Maxwellian plasma rate coefficient $\alpha_{\mathrm{DR}}(T)$ is usually reported using the fitting formula
\begin{equation}
\alpha_{\mathrm{DR}}(T)=\frac{1}{(k_{\mathrm{B}}T)^{3/2}} \sum_{l}{b_{l}\exp{\left(\frac{-E_l}{k_{\mathrm{B}}T}\right)}}
\label{eq:drmaxfit}
\end{equation}
where $b_l$ and $E_l$ are fitting parameters and the total number of terms in the fit is small, typically about ten. This approximation corresponds to a DR cross section that is a series of delta-function resonances at energies $E_{l}$, i.e., $\sigma_{\mathrm{DR}}\propto \sum_{l}{ \sigma_{\mathrm{DR},l} \delta(E - E_{l})}$. The DR rate coefficient for a kappa distribution can then be approximated as \citep{Dzifcakova:SolPhys:1992}
\begin{equation}
\alpha_{\mathrm{DR}}(T_{\kappa}, \kappa) = \frac{A_{\kappa}}{(k_{\mathrm{B}}T_{\kappa})^{3/2}}\sum_{l}{b_{l}
\left[1+\frac{E_{l}}{(\kappa - 3/2)k_{\mathrm{B}T_{\kappa}}}\right]^{-(\kappa+1)}}.
\label{eq:drkapfit}
\end{equation}
\citet{Dzifcakova:SolPhys:1992} shows that this approximation is the first term of a series expansion and so the error can be estimated from the higher order terms. Qualitatively, the uncertainties are due to simplifying the complex resonant structure of the DR cross section into a small number of delta-function resonances. The largest uncertainties are expected for low temperatures, where $E_{l} \sim k_{\mathrm{B}}T$ and for small values of $\kappa \rightarrow 3/2$. 

	Similar methods have been given for approximating other relevant processes, such as radiative recombination (RR) and electron impact ionization \citep[e.g.,][]{Dzifcakova:SolPhys:1992, Wannawichian:ApJS:2003, Dzifcakova:ApJS:2015} and collision strengths for excitation and de-excitation \citep[e.g.,][]{Dudik:AA:2014}. \citet{Dzifcakova:ApJS:2015} recently developed a software package called KAPPA that tabulates kappa distribution rate coefficients for $\kappa = 2, 3, 4, 5, 7, 10, 15, 25,$~and~$33$. This package integrates with the CHIANTI atomic database, which is widely used in the analysis of ultraviolet spectra \citep{Dere:AAS:1997, Landi:ApJS:2013}. Below, we will compare our results using a Maxwellian decomposition approach to results obtained using the KAPPA package.

	The reverse engineering approach has several limitations. First, an adequate approximation must be found for each kind of atomic data, including RR, DR, electron impact ionization, collisional excitation and de-excitation, and so on. Including new atomic processes or improved methods of tabulating Maxwellian rate coefficients requires the development of new approximations. Updates to the Maxwellian data itself can require lengthy recalculations of the kappa distribution rate coefficients. For the case of DR discussed above, the changes are relatively simple substitutions, but for ionization or collisional excitation and de-excitation rates numerical integrations must be performed. Additionally, the approximations can become inaccurate for small $\kappa$. For example, \citet{Dudik:AA:2014} discuss the approximation for distribution average collision strengths, in which the error in the approximation can be 20--30\% for $\kappa = 2$ and grows for smaller values of $\kappa$. Finally, In order to model spectra using the rate coefficients derived in this way, an extensive database of all the required atomic data for each value of $\kappa$ of interest must be constructed. 

\subsection{Maxwellian Decomposition Approach}\label{subsec:summax}

	An alternative approach to deriving atomic data for non-thermal distributions, is to approximate the distribution function itself as a sum of Maxwellians. We refer to this method as the Maxwellian decomposition method. This method has been used in the past. For example, \citet{Ko:GRL:1996} approximated kappa distributions using a sum of Maxwellians in order to model the CSD in the corona and the solar wind. \citet{Kaastra:AA:2009} approximated the distributions of shocks in galaxy clusters using a sum of Maxwellians. 
	
	If a distribution function can be approximated by a sum of Maxwellians, then the rate coefficient for the non-thermal distribution is just a sum of the Maxwellian rate coefficients. Suppose that the arbitrary distribution function $g(E)$ is given by 
\begin{equation}
g(E)=\sum_{j}{c_{j}f(E; T_{j})},
\label{eq:maxapprox}
\end{equation}
where $f(E; T_{j})$ is a Maxwellian at temperature $T_{j}$, and the $c_{j}$ are parameters that weight the sum of Maxwellians. 
From the approximation of Equation~(\ref{eq:maxapprox}) and the expression for the rate coefficient given by Equation~(\ref{eq:ratecoeffdef}), it follows that the rate coefficient for $g(E)$ is
\begin{equation}
\alpha_{\mathrm{g}} = \sum_{j}{c_{j}\alpha(T_{j})},
\label{eq:rateapprox}
\end{equation}
where $\alpha(T_{j})$ is the Maxwellian rate coefficient at the temperature $T_{j}$. Note that there is no requirement that the $c_{j}$ be positive. 

	The Maxwellian decomposition approximation has several potential advantages over the reverse-engineering approach. It is only necessary to calculate the approximation once for a given distribution function. Once the $c_{j}$ and $T_{j}$ are known, they can be applied to every atomic rate coefficient in the same way. Thus, the system is simple, extendable to any atomic process that can be represented as a Maxwellian rate coefficient, and does not need to be updated when new atomic data become available. The approximation for the needed atomic data is, essentially, as good as the approximation of the Maxwellian decomposition of the kappa distribution. Thus, with a suitable decomposition, data for small values of $\kappa$ can be obtained with accuracy as good as that for large values of $\kappa$. Finally, it is straightforward to integrate into existing spectroscopic modeling software. In fact, the CHIANTI database already includes the ability to model spectra for a sum of Maxwellians \citep{Landi:ApJS:2006}. 

\subsection{Method for Finding Kappa Fitting Coefficients}\label{subsec:coeff}

	In order to find fitting parameters that approximate a kappa distribution as a sum of Maxwellians, we have used two numerical procedures. We approximate the kappa distribution using a fit of the form
\begin{equation}
f_{\kappa}(E; \kappa, T_{\kappa}) = \sum_{j}{c_{j}f(E; a_{j}T_{\kappa})}, 
\label{eq:kapapprox}
\end{equation}
where $f(E; a_{j}T_{\kappa})$ is the Maxwellian energy distribution at a temperature of $T=a_{j}T_{\kappa}$. The $a_{j}$ are independent of $T_{\kappa}$, but do depend on $\kappa$ itself. In order to further constrain the $c_{j}$ we impose the normalization condition that
\begin{equation}
\sum_{j}{c_{j}} = 1. 
\label{eq:norm}
\end{equation}
This ensures that the total integral of the distribution function is unity.

	One way to obtain the parameters $c_j$ and $a_j$ is to simultaneously perform a least squares fit to the kappa function using a standard method, such as the Levenberg-Marquardt algorithm \citep{Press:book}. To determine the parameters, a set of energies $E_{i}$ is chosen at which we want to match $f_{\kappa}(E_{i})$. For the solutions given below, we used a linear, evenly spaced set of $E_{i}$ starting at zero energy and extending up to an energy high enough so that less than 0.01\% of the particles have an energy above the maximum $E_{i}$. In order to impose the normalization condition, all of the $c_j$ are constrained to be greater than zero except for one of them, which we label $c_N$, which is set to $1-\sum_{j \neq N}{c_{j}}$ in order to provide the desired normalization constraint. We then performed the least squares fit to all of the parameters. We found that this method works well for moderate to large values of $\kappa \gtrsim 4$. However, for more nonthermal distributions, the fits tend to have a relatively large negative value for $c_{N}$. As discussed below, in Section~\ref{subsec:uncertainties}, in order to reduce the uncertainties in the derived rate coefficients, the negative $c_j$ should be kept small. Thus, we also used a different fitting method, which minimizes any negative $c_j$ values.
	
	This second method for performing the fit uses an iterative approach, in which we find the $c_j$ and the $a_j$ separately. For a given set of $a_{j}$, it is straightforward to find the best fit parameters $c_{j}$. Thus, we begin with an initial guess for the $a_{j}$, solve for the $c_{j}$, then iteratively improve the $a_{j}$ and $c_{j}$ as described below. 
The least squares best approximation to $f_{\kappa}$ is the one that minimizes
\begin{equation}
f_{\kappa}(E_{i}) - \sum_{j}{c_{j}f(E_{i}; a_{j}T_{\kappa})} = r_{i}. 
\label{eq:leastsquares}
\end{equation}
Here $r_{i}$ are the residuals between the Maxwellian decomposition and the kappa distribution. Since the $a_j$ are assumed to be known, and there are usually more energies than terms in the sum of Maxwellians, this is an overdetermined system of linear equations that can be solved using linear least squares, i.e., the solution minimizes $\sum_{i}{r_{i}^2}$. The normalization condition of Equation~(\ref{eq:norm}) can be incorporated as a linear constraint. The best fit values for the $c_{j}$ can then be found using standard methods \citep[e.g.,][]{Meyer:book}. We have used IDL, in which the \textit{la\_least\_square\_equality} routine calculates the solution. There are equivalent functions in the linear algebra packages associated with many other programming languages. 


	It is more difficult to optimize the set of $a_j$. For our intial guess, we start with a few of the $a_j$ below one and the rest evenly spaced between one and the maximum $E_{\mathrm{max}}/k_{\mathrm{B}}T_{\kappa}$. Next the least squares solution for the $c_j$ is found. For the purpose of minimizing the uncertainties (see Section~\ref{subsec:uncertainties}), it is desireable that the magnitudes of the weights $\left|c_j\right|$ be small. This can be accomplished by minimizing the $\sum_{j}{\left|c_j\right|}$. 
	
\begin{figure}
\centering \includegraphics[width=0.9\textwidth]{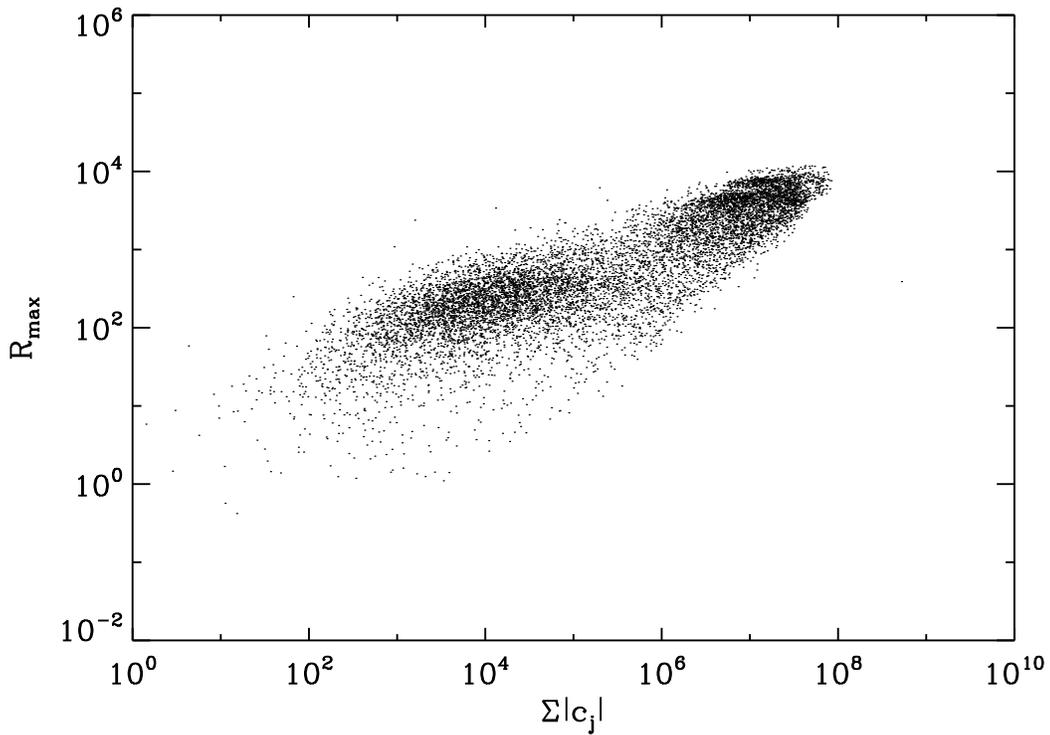}
\caption{\label{fig:totabscorr} The maximum absolute relative error $R_{\mathrm{max}}$ versus the sum of the magnitudes of the $c_j$, $\sum_{j}{\abs{c_{j}}}$.  A random number generator was used to choose the set of $a_j$ for $\kappa = 7$. This shows that there is a strong positive correlation between these quantities, indicating that minimizing $\sum_{j}{\abs{c_{j}}}$ tends to also minimize the error. 
}
\end{figure}

	We have found that $\sum_{j}{\left|c_j\right|}$ is strongly correlated with the residual error between $f_{\kappa}$ and the Maxwellian decomposition. Figure~\ref{fig:totabscorr} shows an example of this correlation for $\kappa = 7$. Various possible sets of $a_{j}$ were selected using a random number generator and the corresponding $c_j$ were found. We then calculated the relative error 
\begin{equation}
R_{\mathrm{err}} \equiv \frac{\abs{f_{\kappa}(E_{i}) - \sum_{j}{c_{j}f(E_i; a_{j}T_{\kappa})}}}{f_{\kappa}}
\label{eq:errdef}
\end{equation}
and plotted the maximum relative error for $E < E_{\mathrm{max}}$
\begin{equation}
R_{\mathrm{max}} = \mathrm{max} \left\{ R_{\mathrm{err}}, E < E_{\mathrm{max}} \right\}
\label{eq:maxerrdef}
\end{equation}
against the $\sum_{j}{\abs{c_j}}$. The Spearman rank correlation coefficient for this plot is 0.86 with at a significance of greater than 99.99\%. Although we are not certain what the underlying reason for this correlation is, we can regard it as an empirical result.
	
	Using this correlation, we can employ a minimization procedure to find the set of $a_j$ that minimizes $\sum_{j}{\abs{c_j}}$. We do this using a downhill simplex method \citep{Press:book}. The result does depend on the initial guess for the $a_j$ and in some cases even after the minimization procedure the results still exceed the desired maximum relative error. In this case, we can use a random number generator to perturb the initial guess and then repeat the procedure until we find a solution that meets our requirements. 
	
	One other factor to consider, using either fitting method, is the number of terms in the sum. We find that a larger number of terms is needed for smaller values of $\kappa$. Our approach to this issue was to start with some moderate number of terms, and then add or remove terms from the sum until it appeared that we had reached the minimum number needed for the desired accuracy.
	
\begin{figure}
\centering \includegraphics[width=0.9\textwidth]{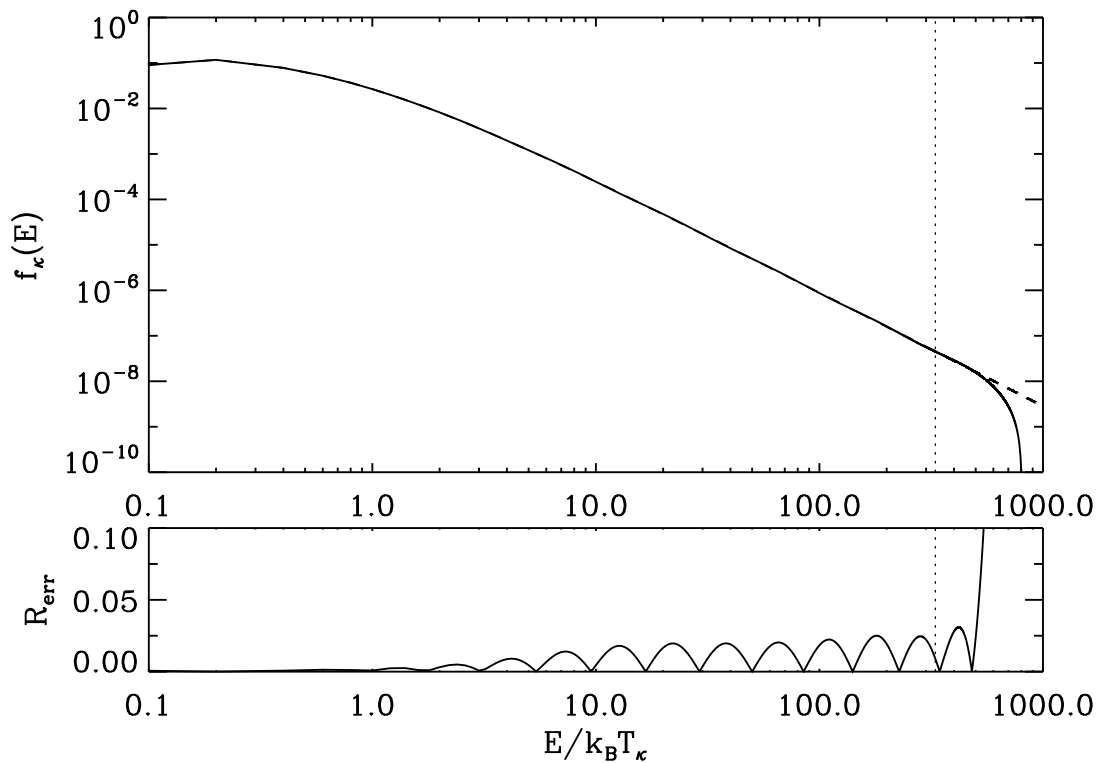}
\caption{\label{fig:kappafit} Example of a Maxwellian decomposition fit to $f_{\kappa}(E)$ for $\kappa = 2$ and $k_{\mathrm{B}}T_{\kappa}=10$ (the units are arbitrary). The dashed curve is the kappa distribution, while the solid curve shows the result obtained by summing Maxwellians. The vertical dotted line indicates the energy at which 99.99\% of the particles have a smaller energy. This corresponds to 330~$k_{\mathrm{B}}T_{\kappa}$ for $\kappa=2$. The lower panel shows $R_{\mathrm{err}}$, which is at most 0.025 for $E < 330$~$k_{\mathrm{B}}T_{\kappa}$. 
}
\end{figure}

	Figure~\ref{fig:kappafit} shows an example of a fit for $\kappa=2$ at $T_{\kappa}=10$. Note that the units of temperature are arbitrary as long as $k_{\mathrm{B}}T_{\kappa}$ has the same units as $E$. The dashed line in the figure shows $f_{\kappa}(E; \kappa, T_{\kappa})$ and the solid line shows our Maxwellian decomposition approximation. The lower panel of the figure illustrates $R_{\mathrm{err}}$, which is below 2.5\% for $E < 330$~$k_{\mathrm{B}}T_{\kappa}$. This energy is indicated by the vertical dotted line on the plot. Fewer than 0.01\% of the particles in the distribution have energies greater than this. 
		
	We have used both of the above fitting methods to find the parameters for the Maxwellian components, and report here the results that were best in terms of having a small error and $\sum_{j}{\abs{c_j}}$. The procedure for finding a set of fitting parameters involves some trial and error. However, once a solution with sufficient precision is found it can be used for all of the atomic data; the procedure only needs to be performed once. A different numerical approach was used by \citet{Kaastra:AA:2009} to approximate an arbitrary distribution function by a sum of 32 Maxwellians. The accuracy of their fits is comparable or sometimes worse than ours. Despite these issues, it is remarkable how well kappa distributions can be approximated by a sum of Maxwellians. We speculate that with a deeper understanding of the mathematical properties of kappa distributions a more systematic approach to this decomposition could be found. 	

\subsection{Uncertainties}\label{subsec:uncertainties}

	The uncertainties in the rate coefficients using the Maxwellian decomposition arise from the goodness of the approximation of the kappa distribution and from the weighting of the Maxwellian terms. This is in contrast to the uncertainties with the reverse engineering method, in which the uncertainties are due to the approximations involved with the individual rate coefficients. Both methods, of course, suffer from the same uncertainties in the Maxwellian atomic data. That is, if the tabulated Maxwellian data is inaccurate, then the derived data for the kappa distribution will also be inaccurate, though not necessarily in exactly the same ways. Since our interest is in comparing the decomposition and reverse-engineering methods, we will omit discussion of errors in the Maxwellian data. 
	
	One source of uncertainty for the Maxwellian decomposition is the accuracy of the approximation to the kappa distribution. This uncertainty can be mitigated by choosing a desired tolerance when finding the fit parameters. Below, in Section~\ref{subsec:coeffdisc}, we describe some cases where we require the maximum error to be less than 3\%. In many cases we were able to obtain even smaller errors. 
	
	Although we characterize the error in the Maxwellian decomposition using the metric $R_{\mathrm{max}}$, the error in the generated kappa-distribution rate coefficients can be different from this value. The error in the rate coefficient is a weighted average over the relevant cross section of the error at all energies, whereas the maximum $R_{\mathrm{err}}$ occurs at the highest energies. If the cross section is largest at low energies and decreasing towards high energies, as is usually the case, the error in the rate coefficient is expected to be smaller than $R_{\mathrm{max}}$. In some cases the cross section is at its maximum at very high energies, such as for some multiple ionization processes. In those cases, $R_{\mathrm{max}}$ may underestimate the error in the rate coefficient. 
	
	An uncertainty also arises due to the truncation of the fit at some maximum energy. That is, the fits are accurate to the desired tolerance for an energy range from $E=0$ to some maximum energy $E_{\mathrm{max}}$. In our typical fits, we chose this energy to be such that no more than 0.01\% of the particles have an energy greater than $E_{\mathrm{max}}$. Because the cross sections for ionization, recombination, and excitation all decrease at sufficiently large energies, this truncation is not expected to produce significant errors in the generated relevant rate coefficients. 
	
	A more complex source of uncertainty comes from the weighting of the terms in the sum. The Maxwellian rate coefficients are functions of temperature, and may have an uncertainty that is also a function of temperature, so that a Maxwellian rate coefficient $\alpha(T_{j})$ has an uncertainty of $\delta_{\alpha}(T_j)$. Assuming they are uncorrelated, the uncertainty in the derived kappa rate coefficient isar
\begin{equation}
\sqrt{ \sum_{j}{ c_j^2 \left[\delta_{\alpha}(a_j T_{\kappa})\right]^2} }. 
\label{eq:sumerr}
\end{equation}
This implies that if any of the magnitudes of the $c_j$ are greater than one, the uncertainty in the kappa-distribution rate coefficient must be greater than that of the Maxwellian rate coefficient from which it is derived, even if the kappa distribution itself is approximated to a very high accuracy. This is the main reason that we minimize $\sum_j{\abs{c_j}}$ in our method. For our fits, none of the $c_j$ is greater than one and the sum does not exceed unity by more than a few percent and usually by less than one percent. The resulting uncertainty in the kappa-distribution rate coefficient is then even smaller, because $\sqrt{\sum_j{c_j^2}} < \sum_j{\abs{c_j}}$. 
	
	Another possible source of uncertainty is related to the accuracy of the tabulated Maxwellian rate coefficients. The decomposition method uses a sum of Maxwellians at various temperatures above and below $T_{\kappa}$ to approximate the kappa distribution. However, the atomic data may be tabulated using a function that is expected to be valid within a certain temperature range; for CHIANTI this is typically 10$^{4}$--10$^{9}$~K. The terms in the sum that fall at lower or higher temperatures may be inaccurate if the Maxwellian data are not accurate at those energies. Because the kappa distribution has a high energy tail, the decomposition is likely to have terms that greatly exceed $T_{\kappa}$, but usually it does not have terms at temperatures so much smaller than the minimum tabulated temperature as to be problematic. The errors at high temperatures are also expected to be small. This is because the $c_j$ at those temperatures are small and because the Maxwellian rate coefficients are usually tabulated using a form that has the correct high energy behavior so that the errors when the rate coefficients are extrapolated to high temperatures are not grossly inaccurate. One caveat, though, is that the tabulated atomic data usually ignore relativistic effects, which may be important for very high temperatures. 

	Finally, because the $c_j$ can be negative it is possible in some situations to find a small negative rate coefficient, which is unphysical. As shown in Section~\ref{subsec:coeffdisc}, the negative values of $c_j$ are usually associated with $a_j \gg 1$. Negative rate coefficients tend to be produced for very low $T_{\kappa}$ in processes that have a threshold, such as ionization. In this case, the largest values of $c_j$ are concentrated at low temperatures where $\alpha(a_{j}T_{\kappa})$ is small, but a negative $c_j$ can occur at a high temperature where the value of $\alpha(a_{j}T_{\kappa})$ is large, resulting in a net negative rate coefficient. In practice, we have found these negative rate coefficients to be insignificant. They only occur in cases where the process is unimportant. For example, with ionization, negative rate coefficients occur at low $T_{\kappa}$ where the abundance of the relevant ion is very small. Thus, any negative rate coefficients that arise can be set to zero with negligible consequences for the CSD or the level populations. In the unlikely case that the negative rate coefficient is problematic, the reader can use the method described in Section~\ref{subsec:coeff} to find an improved Maxwellian decomposition with even smaller $\sum_{j}{\abs{c_j}}$. 
		
\subsection{Coefficients for Certain Values of $\kappa$}\label{subsec:coeffdisc}

	We have found fitting coefficients that describe kappa distributions as a sum of Maxwellians to very high accuracy. Tables~\ref{table:params1}--\ref{table:params5} list the $a_j$ and $c_j$ parameters for $\kappa=1.7$, 2, 3, 4, 5, 7, 10, 15, 20, 25, 30, 33, 50, and 100. The coefficients given in these tables can also be obtained using the IDL code that is included as supplementary online material with this paper. The kappa distribution is approximated by substituting these parameters into Equation~(\ref{eq:kapapprox}). 
	
	Table~\ref{table:paramsacc} gives some parameters describing the accuracy of the Maxwellian decompositions in Tables~\ref{table:params1}--\ref{table:params5}. The accuracy is calculated for $E < E_{\mathrm{max}}$, which is the energy below which 99.99\% of the particles in the kappa distribution are found. The third column of Table~\ref{table:paramsacc} lists the maximum error $R_{\mathrm{max}}$. Also given is the sum of the magnitudes of the $\abs{c_j}$. These are all very close to unity with all of the $c_j < 1$. This implies that the intrinsic errors in the Maxwellian rate coefficients will not be magnified when calculating the kappa distribution rate coefficients. 

\subsection{Coefficients for $1.7 <  \kappa \leq 100$}\label{subsec:coeffcont}

	We have also found that it is possible to find an approximation in terms of a sum of Maxwellians that varies reasonably smoothly as a function of $\kappa$. This allows for the possibility of varying $\kappa$ continuously. In order to find such fits, we fixed the set of $a_j$ values, based on what we found by optimizing the results for particular values of $\kappa$ (see Section~\ref{subsec:coeffdisc} above). We then performed the linear least squares analysis to determine the best fit $c_j$ corresponding to that set of $a_j$. It turns out that the $c_j$ then vary smoothly as a function of $\kappa$, and these $c_j(\kappa)$ can be fit by a polynomial. The resulting fits are accurate within a range of $\kappa$ values, so once the desired accuracy is no longer met, a new set of $a_j$ can be used and the results for the various $c_j$ patched together piecewise. The end result is a continuous approximation of kappa distributions as a Maxwellian decomposition for $\kappa=1.7$--100. The results cannot be expected to be as accurate as the fits found by optimizing the $a_j$ for each $\kappa$. However, the approach enables one to study how properties of spectra vary continuously as a function of $\kappa$, though at the cost of some precision. 
	
\begin{figure}
\centering \includegraphics[width=0.9\textwidth]{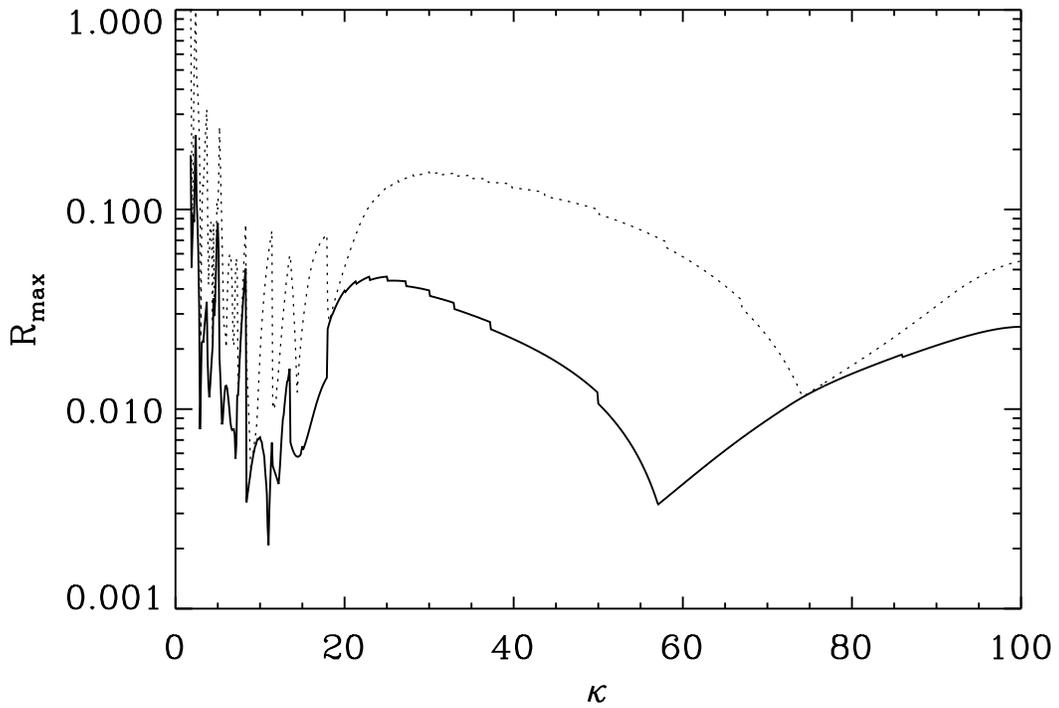}
\caption{\label{fig:polyacc} The maximum relative error $R_{\mathrm{max}}$ between the kappa distribution and the sum of Maxwellians approximated using the values in Table~\ref{table:piecewise}. The solid line indicates the $R_{\mathrm{max}}$ found in the energy range from zero up to the energy that contains 99.9\% of the kappa distribution and the dotted line shows $R_{\mathrm{max}}$ for the higher energy that contains 99.99\% of the particles. 
}
\end{figure}

	 Table~\ref{table:piecewise} gives the $a_j$ and fitting parameters for the $c_j$ in terms of polynomials in $\kappa$, i.e.
\begin{equation}
c_j(\kappa) = \sum_{n}{d_{n}\kappa^{n}}. 
\label{eq:cpolyfit}
\end{equation}
The $a_j$ and $c_j$ reproduced using these polynomial fits can also be obtained by using the IDL code that is included as supplementary online material with this paper. When these $a_j$ and $c_j$ are used to approximate the kappa distribution, the results have $R_{\mathrm{max}} \lesssim 25\%$ in the energy range where 99.9\% of the particles are found. The $R_{\mathrm{max}}$ is typically better than 30\% for the extended energy range that includes 99.99\% of the particles. Figure~\ref{fig:polyacc} illustrates the accuracy as a function of $\kappa$. Given that the uncertainties in the atomic data themselves are typically about 20\%, this is a reasonable accuracy for modeling spectra. Additionally, the magnitudes of the $\abs{c_j}$ sum to less than 1.3 in all cases, and for most $\kappa$ the sum is below 1.02. 
	
	An alternative method for modeling spectra for continuously varying $\kappa$ is to perform a linear interpolation of the rate coefficients between two $\kappa$ values at which the rate coefficients are known accurately. Either method can be used, and the difference between the methods gives an estimate of the uncertainty in the modeled data. We have found small differences of $\lesssim 10\%$ for a few test cases of using linear interpolation of the rate coefficients versus the Maxwellian sum method with the $c_j$ from the polynomial fits in Table~\ref{table:piecewise}. 
	
\section{Comparison of Methods}\label{sec:compare}
	
	As will be shown below, the Maxwellian decomposition and reverse-engineering methods give similar results for generating kappa-distribution atomic data and predicting properties of the plasma, such as the the CSD and line intensity ratios. Summing Maxwellians has, additionally, several other qualitative advantages. A major one is that it can be implemented in a straightforward way using existing spectroscopic analysis software, such as CHIANTI. This is in contrast to reverse-engineering methods, where a new database of rate coefficients must be constructed for each value of $\kappa$. Also, with the decomposition approach, the parameters for the Maxwellian sum can be optimized until an approximation with the desired properties has been obtained. By adding more components to the sum, the approximation can be made accurate to very high energies. As the accuracy of the kappa distribution rate coefficients is mainly determined by the precision with which the Maxwellian sum approximates the kappa distribution, it should be similar for all of the atomic data and any improvements in the decomposition approach are automatically propagated into the derived data. This also implies that the approximation for very small values of $\kappa$ can be made as accurate as that for larger, more Maxwellian, kappa distributions. This is in contrast to the reverse-engineering approach, where the errors increase for small $\kappa$, because some of the approximations used to generate become increasingly inaccurate. Below, we compare our results for the rate coefficients, CSD, and emissivities with those obtained using the reverse engineering approach. Specifically, we use the KAPPA package developed by \citet{Dzifcakova:ApJS:2015}.

\subsection{Rate Coefficients}\label{subsec:rates}

	KAPPA is a database of reverse-engineered rate coefficients and related atomic data derived from those kappa-distribution rate coefficients. \citeauthor{Dzifcakova:ApJS:2015} have, for a few cases, compared of their rate coefficient results with a direct integration of the cross sections. For those test cases they found that the agreement was better than 10\% and in most cases the precision is better than 5\%. In order to give examples from both both relatively smooth and highly resonant processes, we consider ionization and recombination rate coefficients. 
	
\begin{figure}
\centering \includegraphics[width=0.9\textwidth]{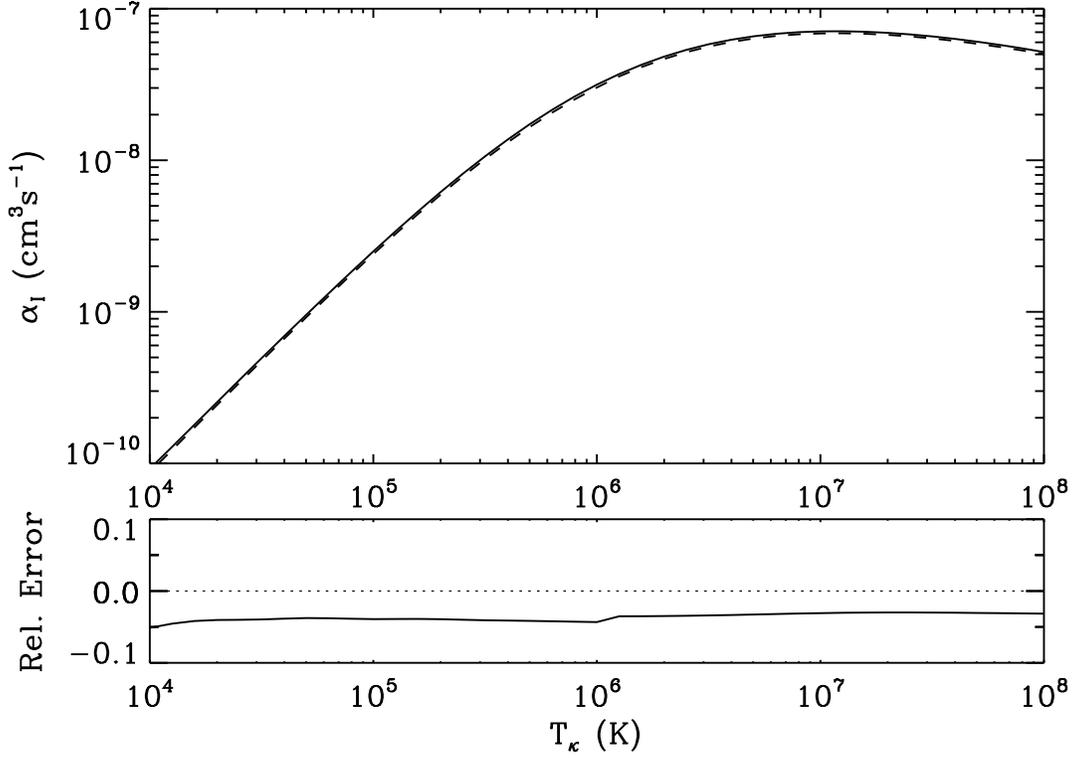}
\caption{\label{fig:ionrate} The ionization rate coefficient $\alpha_{\mathrm{I}}$ for Fe$^{2+}$ forming Fe$^{3+}$ as a function of $T_{\kappa}$ for $\kappa=2$. The solid curve shows our result using the Maxwellian decomposition method, while the dashed line indicates the result using the KAPPA package of \citet{Dzifcakova:ApJS:2015}. The lower panel indicates the relative error given by (KAPPA result -- our result) / our result. 
}
\end{figure}

	For the ionization rate coefficients, we generally find excellent agreement between the reverse-engineering and Maxwellian decomposition approaches. Figure~\ref{fig:ionrate} shows an example of the ionization rate coefficient for Fe$^{2+}$ forming Fe$^{3+}$ with $\kappa=2$. We choose $\kappa=2$ because it is the most nonthermal distribution in the KAPPA package. In the figure our result is shown by the solid curve, while the dashed curve indicates the values given by the KAPPA package. The lower panel shows the relative error (the KAPPA-package result $-$ our result) $/$ our result. In this case we find that the derived rate coefficients agree to better than 5\% at every temperature. We do, however, find that the KAPPA package gives a value that is systematically smaller than ours, by a few percent. Such discrepancies are common for the other kappa rate coefficients and also occurs for the Maxwellian ($\kappa=\infty$) rate coefficient given in that package. For this reason, we suspect that the discrepancy is due to the truncation of the integral over the cross section in the KAPPA package calculations. Additionally, for some of our fits for different $\kappa$, we find discrepancies at low $T_{\kappa}$. This appears to be due to the negative $c_j$ in our approximation. While these can lead to apparently large relative errors, we find that they have negligible influence on the CSD. 
	
\begin{figure}
\centering \includegraphics[width=0.9\textwidth]{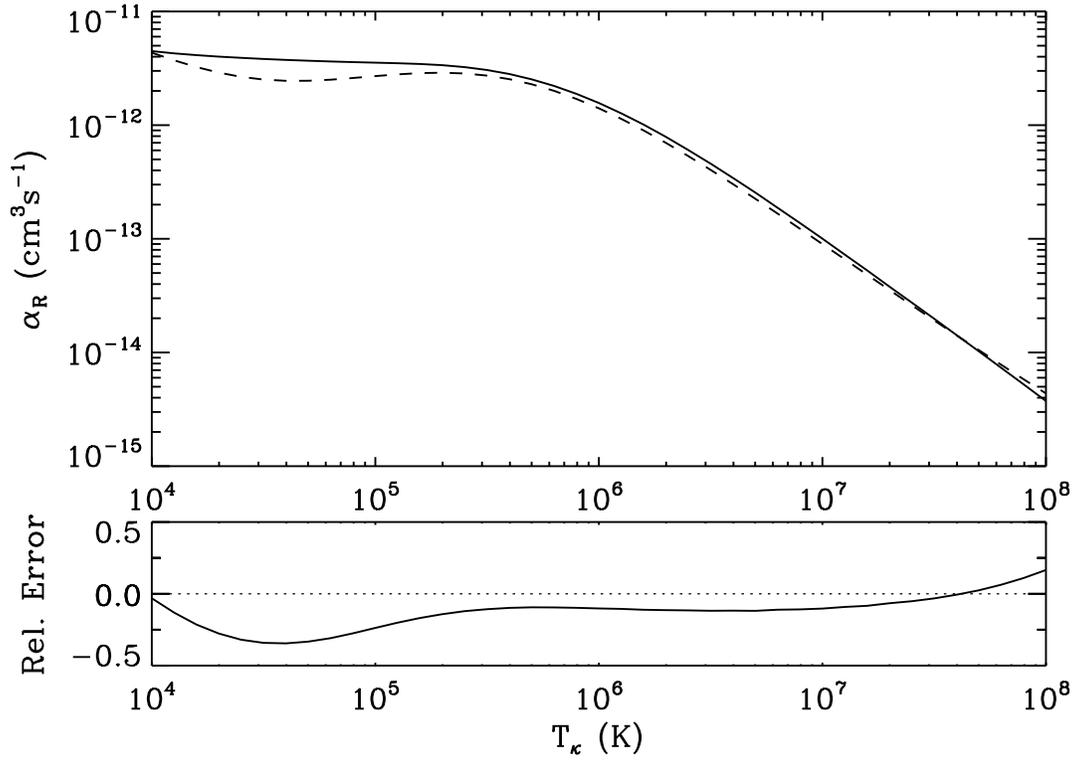}
\caption{\label{fig:recrate} Same as Figure~\ref{fig:ionrate}, but for the total recombination rate coefficient $\alpha_{\mathrm{R}}$ for Fe$^{2+}$ forming Fe$^{1+}$. The rate coefficient includes both RR and DR. 
}
\end{figure}

	The recombination rate coefficients also usually show good agreement, although there are some discrepancies at low $T_{\kappa}$ for low charge states. Figure~\ref{fig:recrate} shows the total recombination rate coefficient (DR + RR) for Fe$^{2+}$ forming Fe$^{1+}$ with $\kappa=2$. The solid curve illustrates our result, which is compared to the result from the KAPPA package as shown by the dashed curve. The lower panel gives the relative difference between the two methods. For this ion we find that there is a discrepancy of about 30\% at low temperatures. A similar discrepancy is present for other values of $\kappa$, which suggests that the error is caused by the approximations used in the reverse-engineering approach. Similar errors are found for other low charged Fe ions, up to about Fe$^{5+}$, but for more highly charged ions the agreement is excellent, within a few percent, over the entire temperature range. For the Fe$^{2+}$ rate coefficient shown in the figure, one can also see a discrepancy at very high temperatures. The reason for this discrepancy is unclear. However, as the abundance of Fe$^{2+}$ is essentially zero at such high energies, this uncertainty will not affect the CSD. 
	
\subsection{Equilibrium Charge State Distribution}\label{subsec:csd}

	The ionization and recombination rate coefficients affect the spectra through the CSD. In order to further compare the Maxwellian decomposition and reverse-engineering, we have calculated the equilibrium CSD using our rate coefficients, ignoring multiple ionization for the moment, and compare to the results given by \citet{Dzifcakova:ApJS:2013} and available in the KAPPA package. 

	The equilibrium CSD is known as collisional ionization equilibrium (CIE). Because the CSD is not evolving in time, the left side of Equation~(\ref{eq:dydt}) is zero. This also implies that the density is a constant factor and plays no role in the solution. For a given temperature, we have a system of algebraic equations. It is easy to see that Equation~(\ref{eq:dydt}) can be written as a matrix 
\begin{equation}
\mathbf{A}\vec{y}=0, 
\label{eq:ay0}
\end{equation}
where $\mathbf{A}$ is the matrix of the rate coefficients and $\vec{y}$ is the vector of abundances with elements $y_{q}$. Equation~(\ref{eq:dydt}) includes multiple ionization rate coefficients, which were not included in the calculations of \citet{Dzifcakova:ApJS:2013}, so for this comparison we will consider only single ionization. The effect of electron impact multiple ionization on the CSD for a kappa distribution is explored in more detail below, in Section~\ref{sec:eimi}. 

In order to obtain a unique solution to Equation~(\ref{eq:ay0}), an additional equation is needed. For this, we require that abundances be normalized so that 
\begin{equation}
\sum_{q}{y_{q}}=1.
\label{eq:norm2}
\end{equation}
This condition is implemented by replacing one of the rows of Equation~(\ref{eq:ay0}) with Equation~(\ref{eq:norm2}), see for example \citet{Bryans:ApJS:2006}. 

\begin{figure}
\centering \includegraphics[width=0.9\textwidth]{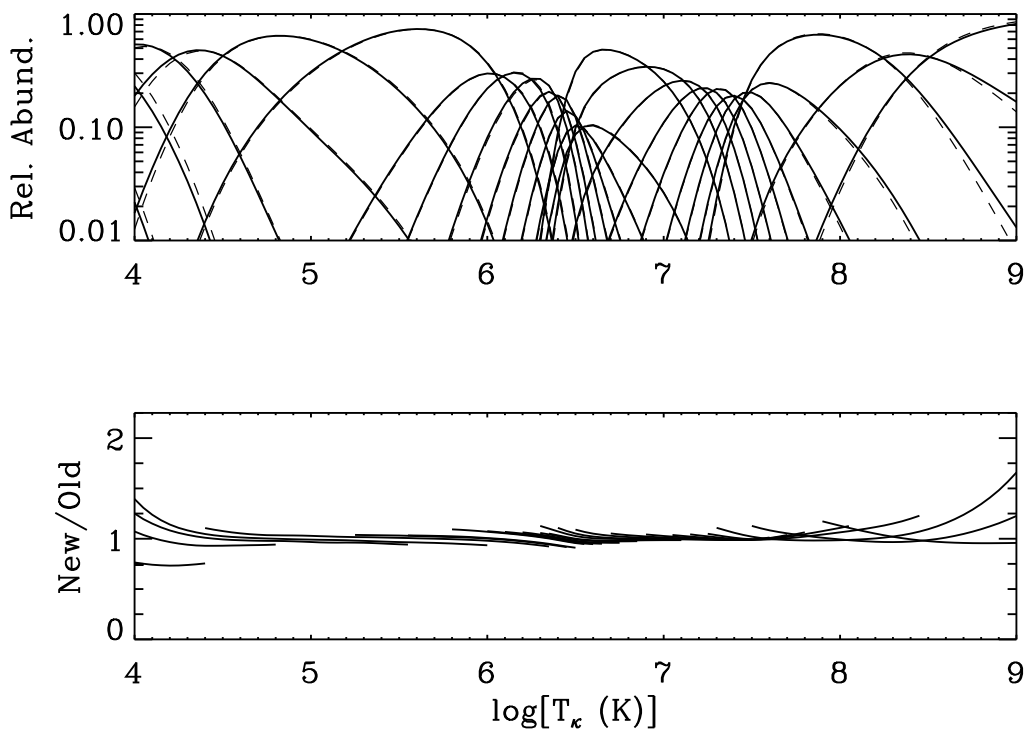}
\caption{\label{fig:csdk2} The CSD for Fe with $\kappa=2$ as a function of $\log{\left[T_{\kappa} (\mathrm{K})\right]}$. The top panel shows the relative abundances of the various charge states, with our results indicated by the solid curves and the abundances given by the KAPPA package shown as dashed curves. The lower panel shows the ratio of our (New) results to those of \citet[][Old]{Dzifcakova:ApJS:2015}. The curves are plotted only for $T_{\kappa}$ where the abundances are greater than 1\%. 
}
\end{figure}

	For temperatures of $T_{\kappa}= 10^{5}$--$10^{8}$~K, we find good agreement between our CSD results and those in the kappa package, with discrepancies of $\lesssim 10\%$. However, at lower and higher temperatures there can be larger differences. The top panel of Figure~\ref{fig:csdk2} shows the relative abundances of Fe charge states for $\kappa=2$, with our results illustrated by the solid curves and those in the KAPPA package indicated by the dashed curves. The lower panel shows the ratio of our (New) calculations to the (Old) calculations of \citet[][]{Dzifcakova:ApJS:2015}. The curves are plotted only for those temperatures where the abundance of the corresponding charge state is greater than 1\%. Over most of the temperature range, the agreement is very good, which is also the case for other values of $\kappa$. However, at very low and very high temperatures, the discrepancies can be $\sim 50\%$. At low temperatures, one cause of the discrepancy is the inaccuracy in the reverse-engineering generated DR rate coefficients. The inaccuracy of the Maxwellian decomposition in the far tail of of the kappa distribution can lead to inaccuracy in the ionization rate coefficients and so can also contribute to discrepancies at low temperatures, as discussed in Section~\ref{subsec:uncertainties}. The reason for the discrepancies in the CSD at high temperatures is not clear. 

\subsection{Level Populations and Line Emissivities}\label{subsec:levelpop}

	In addition to modeling the CSD, the Maxwellian decomposition method can also be used to model the level populations and emissivities of spectral lines. This modeling can be done in a straightforward way using the CHIANTI \textit{emiss\_calc} function with the \textit{sum\_mwl\_coeff} keyword set. This function calculates the emissivities, that is level populations multiplied by radiative decay rates, for a given temperature and density. The keyword, causes the emissivities to be calculated for a distribution that is a sum of Maxwellians weighted by the coefficients $c_{j}$ \citep{Landi:ApJS:2006}. The KAPPA package also has a function for calculating emissivities that is an extension of the CHIANTI function and is called \textit{emiss\_calc\_k} \citep{Dzifcakova:ApJS:2015}.
	
\begin{figure}
\centering \includegraphics[width=0.9\textwidth]{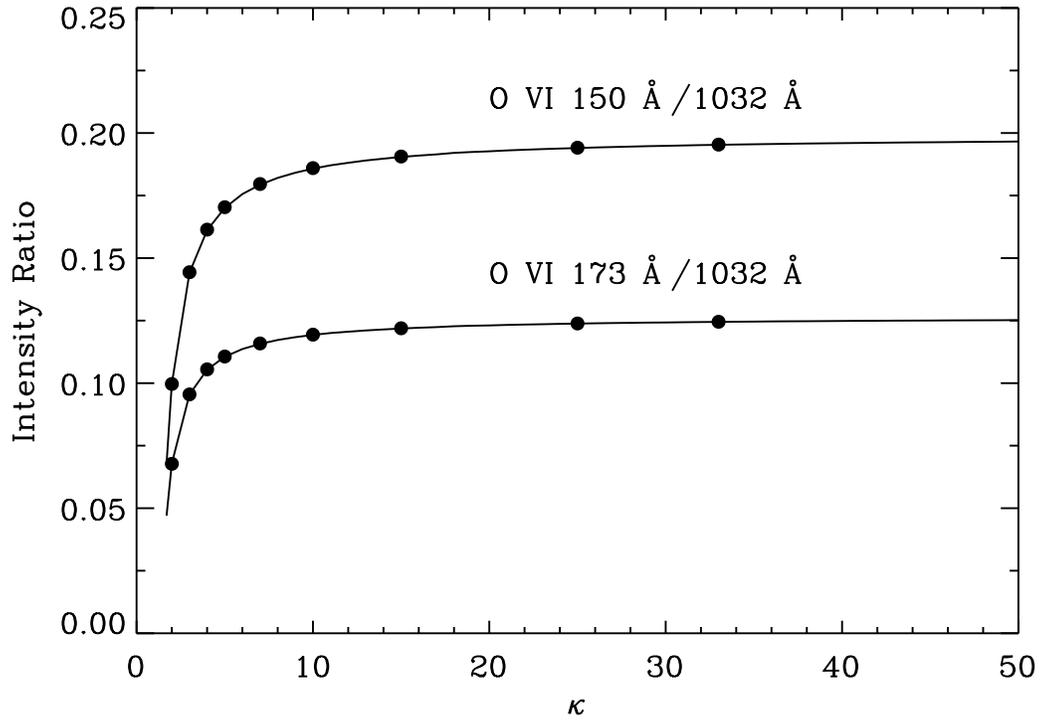}
\caption{\label{fig:intrat} The intensity ratio of two pairs of O~\textsc{vi} lines as a function of $\kappa$. The emissivities were calculated for $T_{\kappa} = 1$~MK and $n_{\mathrm{e}} = 10^{9}$~cm$^{-3}$. The solid curves show our results, which were calculated in a straightforward way using built-in CHIANTI functions (see text for details). The filled circles indicate the ratios obtained using the KAPPA package. 
}
\end{figure}

	Here, we compare our results for the intensity ratio of several transitions of O~\textsc{vi}, which is an abundant ion that is commonly observed in solar physics. The transitions we consider are $1s^2\,2s\;^{2}S_{1/2}\leftarrow 1s^2\,3p\;^{2}P_{1/2,3/2}$ at 150~\AA, $1s^2\,2p\;^{2}P_{1/2,3/2}\leftarrow 1s^2\,3d\;^{2}D_{3/2,5/2}$ at 173~\AA, and $1s^2\,2s\;^{2}S_{1/2} \leftarrow 2s^2\,2p\;^{2}P_{3/2}$ at 1032~\AA. Figure~\ref{fig:intrat} plots the ratios of the 150~\AA/1032~\AA\ and 173~\AA/1032~\AA\ lines as a function of $\kappa$ for a fixed temperature of $T_{\kappa} = 1$~MK and density $n_{\mathrm{e}}=10^{9}$~cm$^{-3}$, which are typical of the solar corona. The solid curves in the figure were obtained using our method, with the Maxwellian distribution parameters found for every value of $\kappa = 1.7$--$50$. The filled circles indicate the ratios given by the KAPPA package the values of $\kappa$ available in the package.
	
	Diagnostics for $\kappa$ based on line intensity ratios have been proposed \citep{Dzifcakova:AA:2011, Dudik:AA:2014, Dudik:ApJ:2015}. One limitation of the reverse-engineering method has been that it is a complicated process to obtain all of the atomic data necessary to calculate these ratios, and so only data for certain discrete values of $\kappa$ are available. The Maxwellian decomposition approach offers the possibility of easily modeling line intensity ratios for any value of $\kappa$. Thus, given a set of observed line ratios, it is possible to find a parameter space of $\kappa$, $T_{\kappa}$ and $n_{\mathrm{e}}$ that is consistent with the data, and constrain all three of these quantities. 
	
\section{Influence of Electron Impact Multiple Ionization on the CSD}\label{sec:eimi}

	Electron impact multiple ionization (EIMI) is a process in which a single electron-ion collision causes two or more bound electrons to be removed from the ion. \citet{Hahn:ApJ:2015} tabulated cross section data for EIMI of Fe, which can be used in CSD calculations. They showed that for Maxwellian plasmas, EIMI has only about a 5\% affect on the CIE abundances, but for dynamic plasmas EIMI can significantly change the evolution of the plasma compared to what is expected if only single ionization is considered in the calculations. 
	
	In plasmas with a kappa EED, the affect of EIMI on the CIE abundances may be much larger than for a Maxwellian plasma. This is because the kappa distribution has a long high energy tail. For a Maxwellian electron distribution, there are few electrons at energies above the energy threshold for EIMI. Thus, for a Maxwellian, very high temperatures are needed for EIMI to occur, but at such high temperatures the abundance of the ion undergoing EIMI is very small. As a result, the net effect of EIMI on the equilibrium CSD is small for a Maxwellian distribution. In contrast, the high energy tail of kappa distributions implies that a significant fraction of nonthermal electrons have energies above the EIMI threshold, even at moderate temperatures. 
	
\begin{figure}
\centering \includegraphics[width=0.9\textwidth]{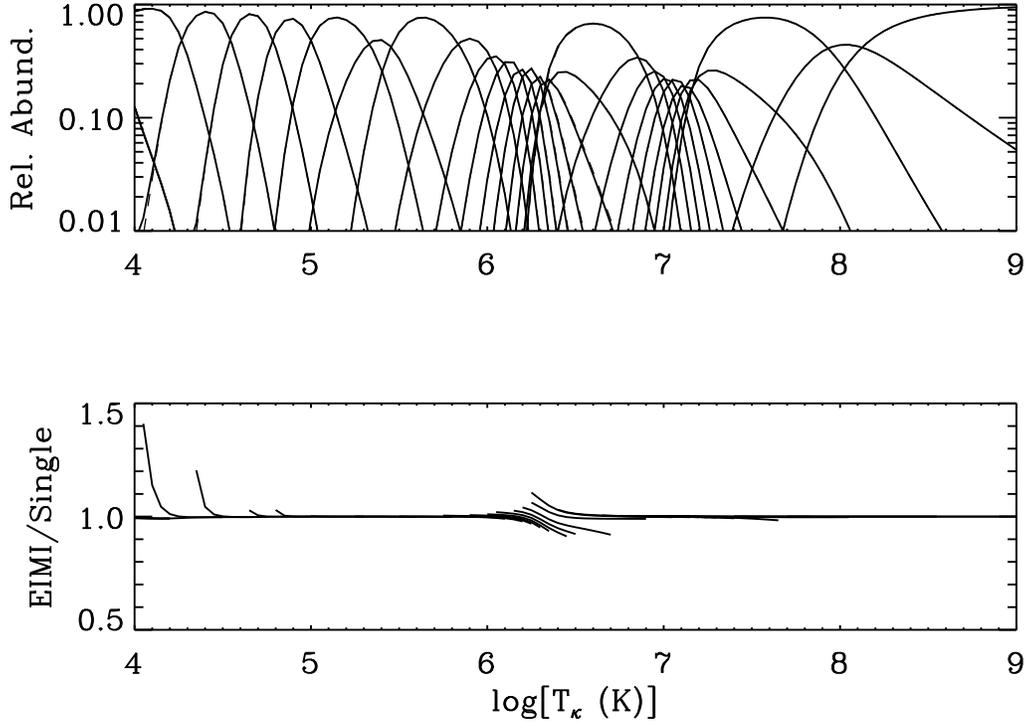}
\caption{\label{fig:eimicsd25} The CSD for Fe as a function of $\log{\left[T_{\kappa} (\mathrm{K})\right]}$ for $\kappa = 25$. The top panel shows the relative abundances of the varous charge states. The solid curve indicates the results when EIMI is included in the calculation and the dashed curve illustrates the CSD when only single ionization is considered. The lower panel shows the ratio of the abundances including (EIMI) or neglecting EIMI (Single). The curves are plotted only for $T_{\kappa}$ where the abundances are greater than 1\%. 
}
\end{figure}

\begin{figure}
\centering \includegraphics[width=0.9\textwidth]{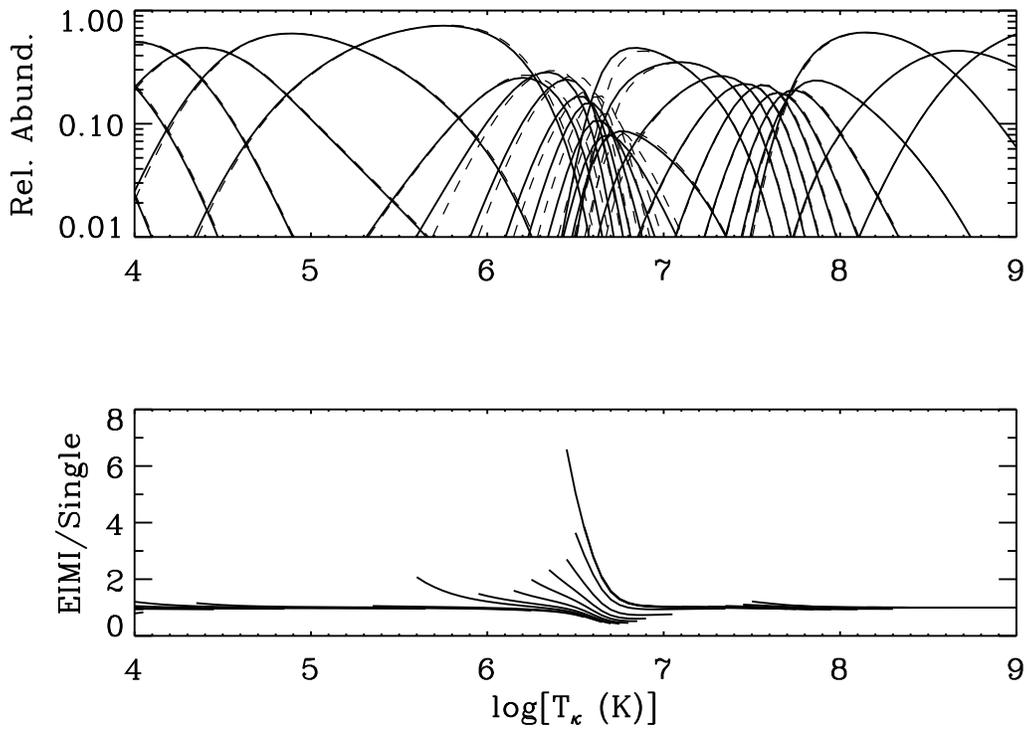}
\caption{\label{fig:eimicsd1p7} Same as Figure~\ref{fig:eimicsd25}, but for $\kappa=1.7$. 
}
\end{figure}

	We have calculated the CIE abundances for iron including EIMI by solving Equation~(\ref{eq:dydt}) with the time derivative set to zero, as described above in Section~\ref{subsec:csd}. However, in this case we keep the multiple ionization terms. The results for a large value of $\kappa= 25$ and a very small $\kappa = 1.7$ are shown in Figures~\ref{fig:eimicsd25} and \ref{fig:eimicsd1p7}, respectively. The solid curve in the figures shows the CIE abundances as a function of $T_{\kappa}$ with EIMI included in the calculation, while the dashed curve illustrates the result if EIMI were ignored. The lower panel shows the ratio of the results that include EIMI to those that use only single ionization. In both plots, the curves are only plotted for $T_{\kappa}$ where the abundances are greater than 1\%. 
	
	The figures show that EIMI has a significant effect on the CSD even in equilibrium for kappa distributions. Even for relatively weak kappa distributions, such as $\kappa=25$ the CSD differs by up to 40\% from what would be expected with only single ionization. In this case, the largest relative effect appears to be in the low charge states, which become ionized to higher charge states at lower temperatures when EIMI is included. The influence of EIMI grows as the distribution becomes more nonthermal. For $\kappa = 1.7$, the largest discrepancies occur for $T_{\kappa}$ between about $10^{6}$ and $10^{7}$~K, where EIMI can change the CSD by a factor of 2 to 7. The iron charge states in this temperature range have their outermost electrons in the $M$ shell. EIMI significantly modifies the CSD at these temperatures, because the kappa distribution contains enough high energy electrons to ionize from the $L$ shell. The resulting $L$-shell hole leads to autoionization of an $M$-shell electron, ending in a net double ionization \citep{Hahn:ApJ:2015}. 

\section{Transient Evolution of a Plasma with Varying $\kappa$}\label{sec:varykap}

\subsection{Timescales}\label{subsec:timescales}

	The availability of rate coefficients for any value of $\kappa$ allows one to model plasmas with a changing $\kappa$. One useful way to quantify the response of plasma to a change in $\kappa$ or $T_{\kappa}$ is to calculate the timescale for the CSD to reach equilibrium following a change in the EED. For Maxwellian EEDs, these timescales are calculated for changes in $T$, and are useful in the analysis of plasmas that evolve following a sudden heating, such as in supernova remnants \citep{Masai:ASS:1984, Hughes:ApJ:1985, Smith:ApJ:2010}. 

	The same methods for calculating these timescales for changes in $T$ in Maxwellian plasmas can be used to derive the timescales for changes in $T_{\kappa}$ and $\kappa$ in kappa plasmas. Following \citet{Masai:ASS:1984}, Equation~(\ref{eq:dydt}) can be written as
\begin{equation}
\frac{d\vec{y}}{dt}=n_{\mathrm{e}}\mathbf{A}(\kappa,T_{\kappa})\vec{y}.
\label{eq:eqtime1}
\end{equation}
If $\kappa$, $T_{\kappa}$ and $n_{\mathrm{e}}$ are constant, such as following a jump from a different set of values for these parameters, then the solution is $\vec{y}(t) = \vec{y}_{0}\exp{\left[n_{\mathrm{e}}t \mathbf{A}(\kappa,T_{\kappa})\right]}$. The exponential of a matrix is defined by the Taylor expansion in powers of the matrix $\mathbf{A}$. The matrix multiplications are more easily carried out by diagonalizing the matrix by finding its eigenvalues and eigenvectors. Doing so, one finds that the solution to Equation~(\ref{eq:eqtime1}) can be written as 
\begin{equation}
\vec{y}(t)=\mathbf{S} \left[ \begin{array}{ccc}
\exp{\left(\lambda_{1}n_{\mathrm{e}}t\right)} & 0 & \cdots \\
0 & \exp{\left(\lambda_{k}n_{\mathrm{e}}t\right)} & 0 \cdots \\
\vdots & 0 & \ddots 
\end{array}\right]\mathbf{S}^{-1}\vec{y}_{0},
\label{eq:eqtime2}
\end{equation}
where $\mathbf{S}$ is the matrix in which each $k$th column is the eigenvector corresponding to the eigenvalue $\lambda_{k}$ of the diagonal matrix. The density-weighted timescales for equilibration are then given by $1/\lambda_{k}$ in units of $n_{\mathrm{e}}t$. Note that the eigenvectors represent linear combinations of the elements of the vector $\vec{y}$, rather than individual elements (i.e., charge states).

\begin{figure}
\centering \includegraphics[width=0.9\textwidth]{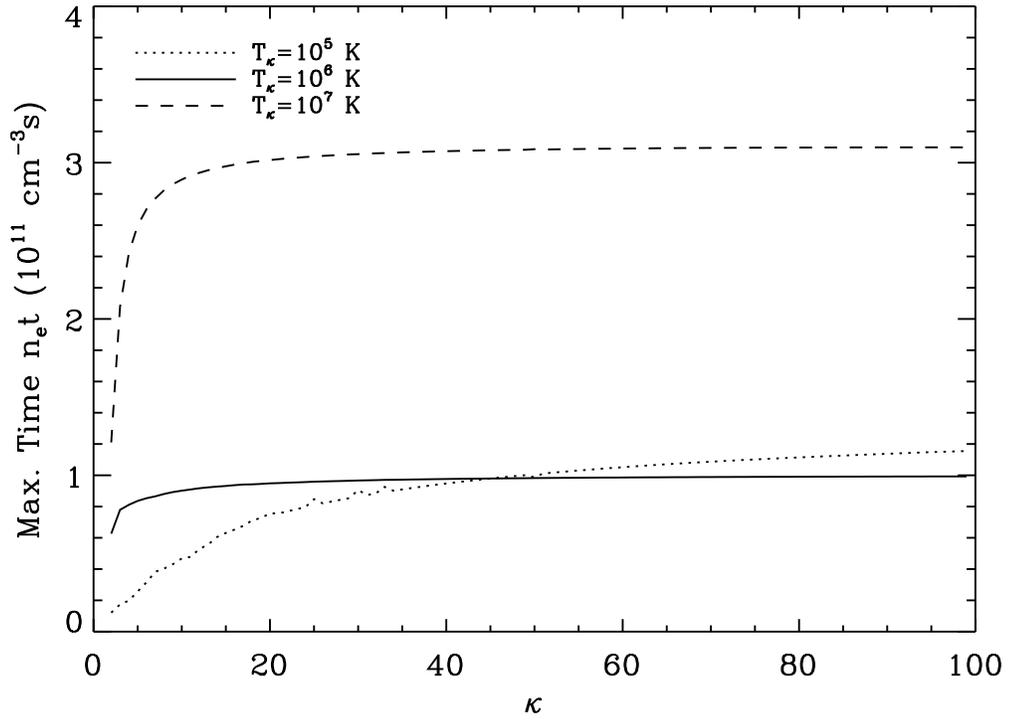}
\caption{\label{fig:maxtime} The equilibration timescale of the CSD as a function of $\kappa$ at several values of $T_{\kappa}$. This shows that the CSD evolves more rapidly for smaller values of $\kappa$, which are the most non-thermal. 
}
\end{figure}

	These timescales could be calculated for any combination of changes in $\kappa$ and $T_{\kappa}$. Here, we show the timescales for a change in $\kappa$, while keeping $T_{\kappa}$ fixed. Figure~\ref{fig:maxtime} shows the maximum timescale $1/\lambda_{k}$ for as a function of $\kappa$ for $T_{\kappa} = 10^{5}$, $10^{6}$, and $10^{7}$~K. This timescale represents the time it would take for the slowest evolving eigenvector to come to equilibrium. The figure shows that the equilibration timescale is shorter for smaller values of $\kappa$, where the distribution is most nonthermal, while the timescale is longer for large $\kappa$ where the distribution is closer to a Maxwellian. An explanation for this is that kappa distributions have both more low energy particles and more high energy particles than a Maxwellian with $T=T_{\kappa}$. This property is becomes more pronounced at smaller values of $\kappa$. Since the recombination rates are generally greater for lower energies and the ionization rates tend to be greater for higher energies, distributions with more high- and low-energy particles respond faster. 
		
\subsection{CSD Evolution for Solar Flares}\label{subsec:flares}

	Nonthermal EEDs are produced in solar flares \citep[e.g.,][]{Seely:ApJ:1987}. Some studies suggest that the energy distributions could be well described by kappa distributions with $\kappa \approx 2$--$8$ \citep{Kasparova:AA:2009, Oka:ApJ:2013}. Using our results for the atomic data for a kappa distribution, we can study how the CSD of the plasma evolves in response to a flare. 
	
	In order to describe the CSD as a function of time, we solve Equation~\ref{eq:dydt} numerically. Initially, we begin with a CSD that corresponds to a Maxwellian distribution, $\kappa=\infty$, with $T_{\kappa} = 1$~MK. This approximately represents the quiescent state of the corona. We assume that the flare turns on as a sudden jump that immediately changes the EED and consequently the rate coefficients, but that the CSD evolves more slowly. For the parameters of the flare EED, we use results from \citet{Kasparova:AA:2009}. They analyzed data from hard X-ray spectra and found that for a coronal source, the spectra could be described using a kappa distribution with $\kappa \approx 8$ and $T_{\kappa} \approx 15$~MK. We used these values in our calculation.
	
\begin{figure}
\centering \includegraphics[width=0.9\textwidth]{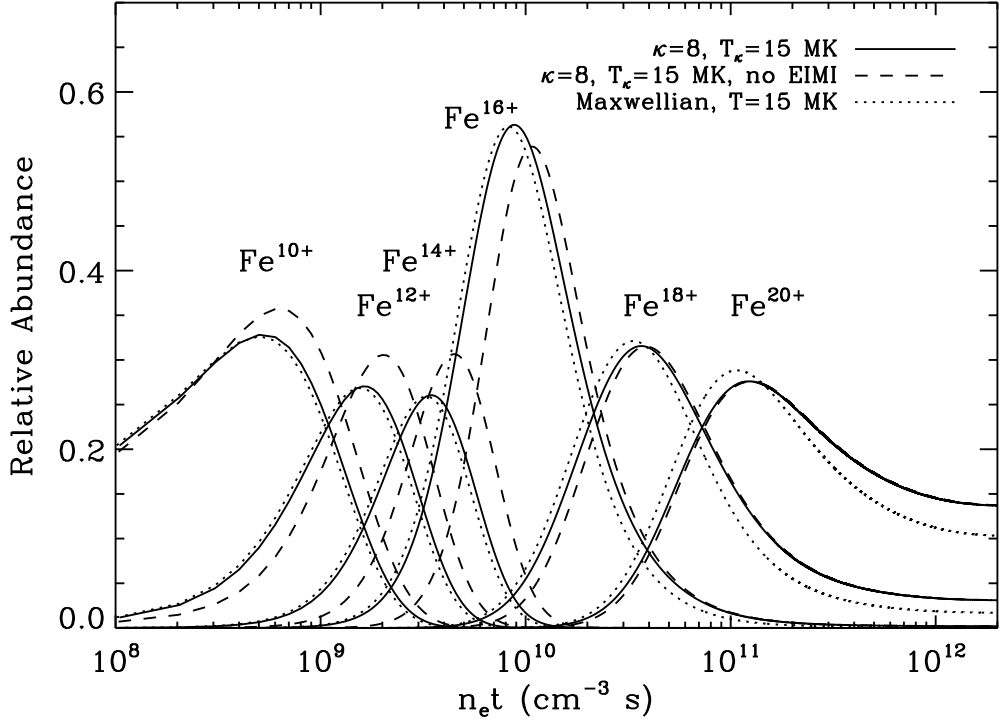}
\caption{\label{fig:corona_nei} The evolution of the CSD in response to a flare, based on parameters inferred by \citet{Kasparova:AA:2009} from hard X-ray observations. The initial condition is a Maxwellian plasma at 1~MK. The solid curve shows the calculated abundances, including EIMI, for a flare that produces electrons with a distribution described by $\kappa=8$ and $T_{\kappa}=15$~MK. The dashed curve illustrates the same conditions, but neglecting EIMI processes. The dotted curve illustrates the results if the plasma were to remain Maxwellian and only have an increased temperature of $15$~MK.
}
\end{figure}

	Figure~\ref{fig:corona_nei} shows the time evolution of several selected charge states of iron, following the sudden change in $\kappa$ and $T_{\kappa}$. Three versions were considered. The solid curves show the case in which EIMI is included, $\kappa=8$, and $T_{\kappa}=15$~MK. The dashed curve illustrates the same scenario, but includes only single ionization. Finally, the dotted curve includes EIMI, but models only the jump in $T_{\kappa}$ and constant $\kappa = \infty$ (i.e., the Maxwellian case). Comparing first the kappa curves with and without EIMI, it is clear that EIMI increases the rate at which the CSD evolves through the lower charge states, but has less of an effect for the higher charge states. Next, comparing the CSD evolution with $\kappa=8$ and EIMI versus the Maxwellian, we find several differences. First, the CSD evolves towards a different equilibrium, as can be seen in the plot at large times where the abundances of Fe$^{18+}$ and Fe$^{20+}$ are asymptoting towards different equilibrium values for the two distributions. The two distributions evolve on similar timescales, although in some cases the $\kappa=8$ CSD lags the Maxwellian CSD. This seems to contradict the timescale analysis above, which implied that the $\kappa$ distribution reaches equilibrium faster than a Maxwellian. However, the timescale analysis considers the general CSD rather than any particular charge state and so is not directly comparable to these results.
For coronal densities of $\sim 10^{8}$--$10^{9}$~$\mathrm{cm^{-3}}$ the difference in the evolving timescales are of order seconds or tens of seconds. Such differences are at about the current temporal resolution of solar observations, such as those from the \textit{Solar Dynamics Observatory} \citep{Lemen:SolPhys:2012, Woods:SolPhys:2012}.
	
	Although we have focused on the parameters relevant to a coronal X-ray source, we have found qualitatively similar results for footpoint sources. The values of $\kappa$ at the footpoints that have been inferred from observations are usually smaller, with values of $\kappa \approx 2$--$3$. The analysis at the footpoints is also more complex as a model of the spectra should include the structure of the chromosphere and transition region, opacity, and hydrodynamic effects in more detail. Nevertheless, our results indicate that the evolution of the plasma CSD in response to a flare can be significantly changed by the presence of a non-thermal distribution, both in the corona or at the footpoints of the flare loop. 
	
\section{Summary}\label{sec:sum}

	We have shown that a kappa distribution can be approximated by a sum of Maxwellians to an accuracy of better than 3\%. This approximation can easily be used to generate collision data needed to model the CSD and other spectroscopic properties of plasmas with kappa EEDs. The resulting data are at least as accurate as those obtained using other methods. In particular, we have found that our results are in good agreement with those obtained using the KAPPA package of \citet{Dzifcakova:ApJS:2015}. This not only demonstrates that our Maxwellian decomposition method is a good alternative to the reverse-engineering approach, but also provides an independent estimate of the uncertainty in that method. For many purposes, either method can be used to a similar level of precision. 
	
	However, summing Maxwellians has some advantages compared to other approaches. These are, first, that it is easy to use and can readily be incorporated into existing spectroscopic analysis software. Second, Maxwellian data for essentially any process can be used to generate kappa-distribution data by using the same small set of fitting parameters. This contrasts with reverse engineering methods where specific approximations must be used for each atomic process and large databases of rate coefficients must be generated for each $\kappa$ in order to model spectra. A related advantage is that any updates in the tabulated Maxwellian atomic data are automatically propagated into kappa distribution data obtained using the decomposition approach. This is in contrast to reverse-engineering methods, for which it is necessary to recalculate the relevant rate coefficients. Additionally, the uncertainties with the decomposition approach mainly reflect the accuracy with which the Maxwellian sum approximates the kappa distribution. Thus, finding a more accurate approximation will improve all of the data. Finally, any value of $\kappa$ can, in principal, be approximated to the same accuracy so that the atomic data derived for small $\kappa$ can be as accurate as that derived for large $\kappa$. 
	
	We have applied the Maxwellian decomposition method to several examples, mainly focusing on the CSD. First, we incorporated EIMI into CIE calcualations and found that EIMI can significantly change the equilibrium CSD. The effect is greatest for small values of $\kappa$, which makes sense as such distributions have the largest nonthermal tails. Next, we studied the evolution of the CSD in a plasma in which both $\kappa$ and $T_{\kappa}$ can vary. For small values of $\kappa$, the CSD is predicted to evolve towards equilibrium faster than those for larger values of $\kappa$ or Maxwellian plasmas. The time dependent CSD equation was solved for several cases that model flares as generating a kappa EED and the time evolution of the various charge states was found. The charge states in a kappa distribution plasma evolve towards a different equilibrium than those for a thermal distribution. EIMI was found to speed up the rate at which the plasma ionizes up through the lower charge states. 
	
	The method of approximating kappa distributions as a sum of Maxwellians is not new, but seems to have fallen out of favor recently in comparison to other methods. This neglect is not justified. The accuracy, ease of use, and portability of the method make it a useful tool for spectroscopic analysis. The same methods can also be applied to other nonthermal distributions beyond kappa distributions. The main issue is to find systematic methods for performing the Maxwellian decomposition to high accuracy.

\begin{acknowledgments}
This work was supported in part by the NASA Living with a Star program grant NNX15AB71G. 
\end{acknowledgments}

\begin{deluxetable}{ccccccc}
\tabletypesize{\small}
\tablecolumns{7}
\tablewidth{0pc}
\tablecaption{Fitting Parameters for Maxwellian Decomposition of Kappa Distributions with $\kappa =$ 1.7, 2, and 3 (Equation~\ref{eq:kapapprox})
\label{table:params1}}
\tablehead{
	\colhead{} &
	\multicolumn{2}{c}{$\kappa=1.7$} & 
	\multicolumn{2}{c}{$\kappa=2$} & 
	\multicolumn{2}{c}{$\kappa=3$} \\
	
	\colhead{$j$} & 
	\colhead{$a_j$} &
	\colhead{$c_j$} & 
	\colhead{$a_j$} &
	\colhead{$c_j$} &
	\colhead{$a_j$} &
	\colhead{$c_j$} 
}
\startdata
0 & 5.00(-2) & 9.5183(-2) & 1.02(-1) & 5.9896(-2) & 1.16(-1) & 2.5922(-3)\\
1 & 1.24(-1) & 3.5911(-1) & 1.68(-1) & 1.3330(-1) & 2.56(-1) & 9.3798(-2)\\
2 & 3.27(-1) & 3.2983(-1) & 3.08(-1) & 3.3109(-1) & 3.54(-1) & 1.0364(-1)\\
3 & 9.23(-1) & 1.4888(-1) & 4.41(-1) & 5.2666(-2) & 5.51(-1) & 3.3185(-1)\\
4 & 2.798(0) & 5.4011(-2) & 7.73(-1) & 2.7887(-1) & 8.72(-1) & 1.8873(-1)\\
5 & 4.266(0) & -9.8173(-3) & 2.043(0) & 1.0852(-1) & 1.391(0) & 1.9666(-1)\\
6 & 7.695(0) & 2.0397(-2) & 5.568(0) & 2.5088(-2) & 2.920(0) & 6.8733(-2)\\
7 & 1.320(1) & -8.5302(-3) & 8.611(0) & 4.1911(-3) & 6.437(0) & 1.1027(-2)\\
8 & 2.0017(1) & 1.6425(-2) & 2.2182(1) & 1.3765(-2) & 1.1726(1) & 2.4131(-3)\\
9 & 2.4556(1) & -8.5330(-3) & 2.6320(1) & -1.1296(-2) & 2.4583(1) & 3.7881(-4)\\
10 & 4.5069(1) & 2.4445(-3) & 3.8883(1) & 3.5490(-3) & 4.1266(1) & 1.9485(-4)\\
11 & 2.22953(2) & 1.1895(-3) & 2.08310(2) & 5.0451(-4) & 1.88810(2) & -4.1575(-5)\\
12 & 3.52864(2) & -1.4590(-3) & 3.11017(2) & -2.2212(-4) & 7.95476(2) & 2.9298(-5)\\
13 & 5.32171(2) & 7.6344(-4) & 9.30643(2) & -1.0136(-4) & \nodata & \nodata \\
14 & 1.27669(3) & 3.0240(-4) & 2.45618(3) & 1.7277(-4) & \nodata & \nodata \\
15 & 3.84582(3) & -1.9796(-4) &  \nodata & \nodata & \nodata & \nodata
\enddata
\tablecomments{The entries of the form a(b) are shorthand for $a\times10^{b}$.}
\end{deluxetable}

\begin{deluxetable}{ccccccc}
\tabletypesize{\small}
\tablecolumns{7}
\tablewidth{0pc}
\tablecaption{Same as Table~\ref{table:params1} but for $\kappa = $ 4, 5, and 7
\label{table:params2}}
\tablehead{
	\colhead{} & 
	\multicolumn{2}{c}{$\kappa=4$} & 
	\multicolumn{2}{c}{$\kappa=5$} & 
	\multicolumn{2}{c}{$\kappa=7$} \\
	
	\colhead{$j$} & 
	\colhead{$a_j$} &
	\colhead{$c_j$} & 
	\colhead{$a_j$} &
	\colhead{$c_j$} &
	\colhead{$a_j$} &
	\colhead{$c_j$} 
}
\startdata
0 & 3.29(-1) & 1.1589(-1) & 4.62(-1) & 2.6404(-1) & 4.92(-1) & 1.6817(-1) \\
1 & 5.98(-1) & 3.9529(-1) & 8.64(-1) & 4.9170(-1) & 8.12(-1) & 4.8515(-1) \\
2 & 1.053(0) & 3.2316(-1) & 1.577(0) & 2.0295(-1) & 1.320(0) & 2.8210(-1) \\
3 & 1.844(0) & 1.2715(-1) & 2.921(0) & 3.7629(-2) & 2.195(0) & 5.9684(-2) \\
4 & 3.302(0) & 3.2476(-2) & 5.936(0) & 3.5942(-3) & 3.989(0) & 4.8443(-3) \\
5 & 6.411(0) & 5.5656(-3) & 2.0618(1) & 8.1131(-5) & 1.1406(1) & 5.3502(-5) \\
6 & 1.5520(1) & 4.7317(-4) & \nodata & \nodata & \nodata & \nodata 
\enddata
\end{deluxetable}

\begin{deluxetable}{ccccccc}
\tabletypesize{\small}
\tablecolumns{7}
\tablewidth{0pc}
\tablecaption{Same as Table~\ref{table:params1} but for $\kappa = $ 10, 15, and 20
\label{table:params3}}
\tablehead{
	\colhead{} & 
	\multicolumn{2}{c}{$\kappa=10$} & 
	\multicolumn{2}{c}{$\kappa=15$} & 
	\multicolumn{2}{c}{$\kappa=20$} \\
	
	\colhead{$j$} & 
	\colhead{$a_j$} &
	\colhead{$c_j$} & 
	\colhead{$a_j$} &
	\colhead{$c_j$} &
	\colhead{$a_j$} &
	\colhead{$c_j$} 
}
\startdata
0 & 5.18(-1) & 9.5556(-2) & 6.37(-1) & 1.9206(-1) & 6.06(-1) & 5.9165(-2)\\
1 & 7.84(-1) & 4.5424(-1) & 9.61(-1) & 6.3165(-1) & 8.30(-1) & 4.4999(-1)\\
2 & 1.190(0) & 3.7406(-1) & 1.526(0) & 1.7558(-1) & 1.139(0) & 4.2849(-1)\\
3 & 1.920(0) & 7.4674(-2) & 3.966(0) & 7.0226(-4) & 1.643(0) & 6.2175(-2)\\
4 & 5.058(0) & 4.8765(-3) & 5.633(0) & 1.3855(-5) & 3.475(0) & 1.7974(-4)\\
5 & 5.550(0) & -3.3968(-3) & 7.0359(1) & -8.8205(-6) & \nodata & \nodata
\enddata
\end{deluxetable}

\begin{deluxetable}{ccccccc}
\tabletypesize{\small}
\tablecolumns{7}
\tablewidth{0pc}
\tablecaption{Same as Table~\ref{table:params1} but for $\kappa = $ 25, 30, and 33
\label{table:params4}}
\tablehead{
	\colhead{} & 
	\multicolumn{2}{c}{$\kappa=25$} & 
	\multicolumn{2}{c}{$\kappa=30$} & 
	\multicolumn{2}{c}{$\kappa=33$} \\
	
	\colhead{$j$} & 
	\colhead{$a_j$} &
	\colhead{$c_j$} & 
	\colhead{$a_j$} &
	\colhead{$c_j$} &
	\colhead{$a_j$} &
	\colhead{$c_j$} 
}
\startdata
0 & 6.62(-1) & 1.0281(-1) & 7.29(-1) & 1.8056(-1) & 7.16(-1) & 1.2455(-1)\\
1 & 9.26(-1) & 6.486(-1) & 9.76(-1) & 6.3110(-1) & 9.39(-1) & 5.8530(-1)\\
2 & 1.328(0) & 2.4775(-1) & 1.333(0) & 1.8601(-1) & 1.185(0) & 2.1208(-1)\\
3 & 2.827(0) & 8.4738(-4) & 1.873(0) & 2.2955(-3) & 1.406(0) & 7.8024(-2) \\
4 & 1.4319(1) & -1.5625(-5) & 8.438(0) & 5.3631(-5) & 6.374(0) & 5.0427(-5) \\
5 & 6.9158(1) & 3.4698(-6) & 3.145(1) & -1.8321(-5) & 8.7117(1) & -7.4012(-6)
\enddata
\end{deluxetable}

\begin{deluxetable}{ccccc}
\tabletypesize{\small}
\tablecolumns{5}
\tablewidth{0pc}
\tablecaption{Same as Table~\ref{table:params1} but for $\kappa = $ 50 and 100
\label{table:params5}}
\tablehead{
	\colhead{} & 
	\multicolumn{2}{c}{$\kappa=50$} & 
	\multicolumn{2}{c}{$\kappa=100$} \\ 
	
	\colhead{$j$} & 
	\colhead{$a_j$} &
	\colhead{$c_j$} & 
	\colhead{$a_j$} &
	\colhead{$c_j$} 
}
\startdata
0 & 7.23(-1) & 5.0910(-2) & 8.43(-1) & 2.1409(-1)\\
1 & 9.16(-1) & 6.0722(-1) & 1.023(0) & 7.2273(-1)\\
2 & 1.190(0) & 3.4184(-1) & 1.273(0) & 6.3198(-2)\\
3 & 1.2300(1) & 4.6799(-5) & 2.0710(1) & -2.1750(-5) \\
4 & 1.00200(2)& -1.6831(-5)& \nodata & \nodata 
\enddata
\end{deluxetable}

\begin{deluxetable}{cccc}
\tabletypesize{\small}
\tablecolumns{4}
\tablewidth{0pc}
\tablecaption{Accuracies of Fitting Parameters in Tables~\ref{table:params1}--\ref{table:params5}
\label{table:paramsacc}}
\tablehead{
	\colhead{$\kappa$} & 
	\colhead{$E_{\mathrm{max}}/k_{\mathrm{B}}T_{\kappa}$\tablenotemark{a}} & 
	\colhead{$R_{\mathrm{max}}$ for $E < E_{\mathrm{max}}$} & 
	\colhead{$\sum_{j}{\abs{c_j}}$}		
}
\startdata
1.7 & 632 & 0.0214 & 1.057 \\
2 & 330 & 0.0250 & 1.023 \\
3 & 77.7 & 0.0194 & 1.000 \\
4 & 41.5 & 0.0020 & 1.000 \\
5 & 29.7 & 0.0032 & 1.000 \\
7 & 21.1 & 0.0009 & 1.000 \\
10 & 16.7 & 0.0032 & 1.007 \\
15 & 14.2 & 0.0262 & 1.000 \\
20 & 13.1 & 0.0017 & 1.000 \\
25 & 12.5 & 0.0165 & 1.000 \\
30 & 12.2 & 0.0077 & 1.000 \\
33 & 12.0 & 0.0055 & 1.000 \\
50 & 11.5 & 0.0282 & 1.000 \\
100 & 11.0 & 0.0020 & 1.000
\enddata
\tablenotetext{a}{$E_{\mathrm{max}}$ is the energy below which 99.99\% of the particles are found, i.e., $\int_{0}^{E_{\mathrm{max}}}{f_{\kappa}(E; \kappa, T_{\kappa}) \diff{E}} = 0.9999$.}
\end{deluxetable}

\begin{deluxetable}{ccccccccc}
\tabletypesize{\scriptsize}
\tablecolumns{9}
\tablewidth{0pc}
\tablecaption{Fitting Kappa Distributions for Continuously Varying $\kappa = $ 1.7 -- 100
\label{table:piecewise}}
\tablehead{
	\colhead{Range} & 
	\colhead{$a_j$} & 
	\colhead{$d_0$} & 
	\colhead{$d_1$} & 
	\colhead{$d_2$} & 
	\colhead{$d_3$} & 
	\colhead{$d_4$} & 
	\colhead{$d_5$} & 
	\colhead{$d_6$}
}
\startdata
$1.7 < \kappa \leq 2.0$ & 0.049 & 1.56729(0) & -1.59895(0) & 4.07812(-1) & & & & \\
\ditto & 0.080 & 8.35027(0) & -1.08223(1) & 4.68258(0) & -6.76624(-1) & & & \\
\ditto & 0.168 & -3.94995(-1) & 2.00317(0) & -1.33084(0) & 2.42976(-1) & & & \\
\ditto & 0.384 & -7.98933(0) & 1.06867(1) & -4.51282(0) & 6.33092(-1) & & & \\
\ditto & 0.944 & 1.03347(1) & -1.42760(1) & 6.50682(0) & -9.48923(-1) & & & \\
\ditto & 1.118 & -3.37385(1) & 5.75809(1) & -3.66138(1) & 1.03267(1) & -1.09760(0) & & \\
\ditto & 2.267 & 2.54497(0) & -3.50125(0) & 1.59975(0) & -2.34561(-1) & & & \\
\ditto & 4.362 & -1.92314(0) & 2.58894(0) & -1.12945(0) & 1.61818(-1) & & & \\
\ditto & 12.253 & 3.98211(-1) & -5.25104(-1) & 2.35482(-1) & -3.49376(-2) & & & \\
\ditto & 38.992 & -3.94167(0) & 7.29473(0) & -5.04540(0) & 1.54697(0) & -1.77514(-1) & & \\
\ditto & 153.710 & 1.56669(0) & -1.54177(0) & 3.79314(-1) & & & & \\
\ditto & 199.674 & -2.03669(0) & 1.99049(0) & -4.86039(-1) & & & & \\
\ditto & 393.810 & 7.76167(-1) & -7.39125(-1) & 1.75439(-1) & & & & \\
\ditto & 997.366 & -1.76300(-1) & 1.61646(-1) & -3.67022(-2) & & & & \\

$2.0 < \kappa \leq 2.3$ & 0.049 & -4.77764(-2) & 4.21978(-2) & -9.00028(-3) & & & & \\
\ditto & 0.080 & 8.35027(0) & -1.08223(1) & 4.68258(0) & -6.76624(-1) & & & \\
\ditto & 0.168 & -3.94995(-1) & 2.00317(0) & -1.33084(0) & 2.42976(-1) & & & \\
\ditto & 0.384 & -7.98933(0) & 1.06867(1) & -4.51282(0) & 6.33092(-1) & & & \\
\ditto & 0.944 & 1.03347(1) & -1.42760(1) & 6.50682(0) & -9.48923(-1) & & & \\
\ditto & 1.118 & -3.37385(1) & 5.75809(1) & -3.66138(1) & 1.03267(1) & -1.09760(0) & & \\
\ditto & 2.267 & 2.54497(0) & -3.50125(0) & 1.59975(0) & -2.34561(-1) & & & \\
\ditto & 4.362 & -1.92314(0) & 2.58894(0) & -1.12945(0) & 1.61818(-1) & & & \\
\ditto & 12.253 & 3.98211(-1) & -5.25104(-1) & 2.35482(-1) & -3.49376(-2) & & & \\
\ditto & 38.992 & -3.94167(0) & 7.29473(0) & -5.04540(0) & 1.54697(0) & -0.177514(-1) & & \\
\ditto & 153.710 & 3.13273(-2) & -2.96564(-2) & 7.09938(-3) & & & & \\
$2.3 < \kappa < 2.4$ & 0.049 & -4.77764(-2)& 4.21978(-2) & -9.00028(-3) & & & & \\
\ditto & 0.080 & 8.35027(0) & -1.08223(1) & 4.68258(0) & -6.76624(-1) & & & \\
\ditto & 0.168 & -3.94995(-1) & 2.00317(0) & -1.33084(0) & 2.42976(-1) & & & \\
\ditto & 0.384 & -7.98933(0) & 1.06867(1) & -4.51282(0) & 6.33092(-1) & & & \\
\ditto & 0.944 & 1.03347(1) & -1.42760(1) & 6.50682(0) & -9.48923(-1) & & & \\
\ditto & 1.118 & -3.37385(1) & 5.75809(1) & -3.66138(1) & 1.03267(1) & -1.09760(0) & & \\
\ditto & 2.267 & 2.54497(0) & -3.50125(0) & 1.59975(0) & -2.34561(-1) & & & \\
\ditto & 4.362 & -1.92314(0) & 2.58894(0) & -1.12945(0) & 1.61818(-1) & & & \\
\ditto & 12.253 & 3.98211(-1) & -5.25104(-1) & 2.35482(-1) & -3.49376(-2) & & & \\
\ditto & 38.992 & -3.94167(0) & 7.29473(0) & -5.04540(0) & 1.54697(0) & -1.77514(-1) & & \\
$2.4 \leq \kappa < 3.7$ & 0.116 & 3.96920(0) & -5.61929(0) & 3.19670(0) & -9.11843(-1) & 1.30249(-1) & -7.44621(-3) & \\
\ditto & 0.256 & 5.64248(0) & -5.33407(0) & 1.93157(0) & -3.12193(-1) & 1.78935(-2) & 2.02901(-4) & \\
\ditto & 0.354 & -8.32819(0) & 9.51261(0) & -4.18395(0) & 8.93371(-1) & -9.16849(-2) & 3.51840(-3) & \\
\ditto & 0.551 & 3.37016(0) & -2.76432(0) & 7.98392(-1) & -3.48088(-2) & -1.90868(-2) & 2.28428(-3) & \\
\ditto & 0.872 & -3.64079(0) & 3.60739(0) & -1.27424(0) & 1.93318(-1) & -7.60047(-3) & -5.29236(-4) & \\
\ditto & 1.391 & 9.04059(-1) & -6.29854(-1) & 1.24991(-1) & 2.96435(-2) & -1.25896(-2) & 1.13885(-3) & \\
\ditto & 2.920 & -1.04513(0) & 1.22861(0) & -5.33511(-1) & 1.14042(-1) & -1.19138(-2) & 4.74984(-4) & \\
\ditto & 6.437 & 3.96190(-1) & -3.53759(-1) & 1.23179(-1) & -1.88395(-2) & 7.98385(-4) & 4.81100(-5) & \\
\ditto & 11.726 & -4.96928(-1) & 6.30421(-1) & -3.20607(-1) & 8.18266(-2) & -1.04245(-2) & 5.28095(-4) & \\
\ditto & 24.583 & 2.28994(-1) & -2.77800(-1) & 1.37518(-1) & -3.45280(-2) & 4.36068(-3) & -2.20187(-4) & \\
$3.7 \leq \kappa < 4.5$ & 0.302 & 1.96013(0) & -1.10591(0) & 2.18455(-1) & -1.48857(-2) & & & \\
\ditto & 0.561 & -2.15731(0) & 1.69698(0) & -3.70337(-1) & 2.65640(-2) & & & \\
\ditto & 1.043 & 3.80733(0) & -2.45503(0) & 5.63680(-1) & -4.15712(-2) & & & \\
\ditto & 1.590 & -4.73982(0) & 3.19782(0) & -7.01062(-1) & 5.08460(-2) & & & \\
\ditto & 2.582 & 2.76133(0) & -1.77615(0) & 3.89392(-1) & -2.83457(-2) & & & \\
\ditto & 6.215 & -1.73113(0) & 1.17320(0) & -2.60640(-1) & 1.90386(-2) & & & \\
\ditto & 9.626 & 1.43348(0) & -9.54201(-1) & 2.09719(-1) & -1.52106(-2) & & & \\
\ditto & 17.292& 3.34049(-1) & 2.23321(-1) & -4.91735(-2) & 3.56517(-3) & & & \\
$4.5 \leq \kappa < 5.2$ & 0.302 & 6.04673(0) & -5.53465 (0) & 2.11477(0) & -4.14777(-1) & 4.136333(-2) & -1.66875(-3) & \\
\ditto & 0.561 & -5.59121(0) & 4.95522(0) & -1.52890(0) & 2.09511(-1) & -1.08252(-2) & & \\
\ditto & 1.043 & 1.34115(1) & -1.13733(1) & 3.66504(0) & -5.20241(-1) & 2.76641(-2) & & \\
\ditto & 1.590 & -2.08424(1) & 1.79968(1) & -5.79182(0) & 8.2769(-1) & -4.43552(-2) & & \\
\ditto & 2.528 & 1.72935(1) & -1.49928(1) & 4.88582(0) & -7.06458(-1) & 3.82436(-2) & & \\
\ditto & 6.215 & -2.53090(1) & 2.23218(1) & -7.35343(0) & 1.07306(0) & -5.85483(-2) & & \\
$5.2 \leq \kappa < 7.3$ & 0.397 & 1.69355(0) & -7.38559(-1) & 1.29662(-1) & -1.05755(-2) & 3.32008(-4) & & \\
\ditto & 0.662 & -1.64244(0) & 1.04587(0) & -1.98344(-1) & 1.67171(-2) & -5.32776(-4) & & \\
\ditto & 1.130 & 1.36114(0) & -6.07907(-1) & 1.29423(-1) & -1.12748(-2) & 3.60750(-4) & & \\
\ditto & 2.133 & -7.80854(-1) & 4.69366(-1) & -8.92911(-2) & 7.22596(-3) & -2.15370(-4) & & \\
\ditto & 6.708 & -8.22231(-1) & 8.95811(-1) & -3.17094(-1) & 5.10962(-2) & -3.93547(-3) & 1.19768(-4) & \\
\ditto & 10.978& 7.02190(-1) & -6.65658(-1) & 1.87408(-1) & -2.06380(-2) & 7.72861(-4) & & \\
\ditto & 19.325& -3.57098(0) & 2.81704(0) & -8.50939(-1) & 1.25010(-1) & -9.07755(-2) & 2.65483(-4) & \\
\ditto & 49.952& 4.506(-1) & -3.080(-1) & 7.376(-2) & -7.380(-3) & 2.596(-4) & & \\
$7.3 \leq \kappa < 8.4$ & 0.397 & 1.69355(0) & -7.38559(-1) & 1.29662(-1) & -1.05755(-2) & 3.32008(-4) & & \\
\ditto & 0.662 & -1.64244(0) & 1.04587(0) & -1.98344(-1) & 1.67171(-2) & -5.32776(-4) & & \\
\ditto & 1.130 & 1.36114(0) & -6.07907(-1) & 1.29423(-1) & -1.12748(-2) & 3.60750(-4) & & \\
\ditto & 2.133 & -7.80854(-1) & 4.69366(-1) & -8.92911(-2) & 7.22596(-3) & -2.15370(-4) & & \\
\ditto & 6.708 & 1.18523(-1) & -4.04953(-2) & 4.58953(-3) & -1.68968(-4) & & & \\
\ditto & 10.978& -4.96980(-2) & 1.78065(-2) & -2.08221(-3) & 7.77779(-5) & & & \\
$8.4 \leq \kappa < 11.5$ & 0.452 & 1.30350(0) & -4.49713(-1) & 6.88978(-2) & -5.82877(-3) & 2.81686(-4) & -7.30064(-6) & 7.88427(-8) \\
\ditto & 0.694 & -1.66061(0) & 8.73637(-1) & -1.52021(-1) & 1.38576(-2) & -7.05050(-4) & 1.89861(-5) & -2.11219(-7) \\ 
\ditto & 1.074 & 2.24466(0) & -9.18802(-1) & 1.76499(-1) & -1.69900(-2) & 8.96474(-4) & -2.47898(-5) & 2.81438(-7) \\
\ditto & 1.791 & -1.56419(0) & 7.86300(-1) & -1.45517(-1) & 1.38959(-2) & -7.32919(-4) & 2.03092(-5) & -2.31248(-7) \\
\ditto & 5.315 & 4.89878(-1) & -1.58942(-1) & 2.12165(-2) & -1.534327(-3) & 6.24996(-5) & -1.10994(-6) & \\
\ditto & 9.741 & -3.01963(-1) & 6.00271(-2) & -2.08105(-3) & -1.66306(-4) & 7.59614(-6) & & \\
\ditto & 22.641& 1.01885(-1) & 1.06359(-2) & -7.32292(-3) & 7.19906(-4) & -1.96401(-5) & & \\
\ditto & 74.007& -4.89289(-3) & -1.91647(-2) & 5.18355(-3) & -4.33200(-4) & 1.11707(-5) & & \\
$11.5 \leq \kappa \leq 13.5$ & 0.452 & 1.30350(0) & -4.49713(-1) & 6.88978(-2) & -5.82877(-3) & 2.81686(-4) & -7.30064(-6) & 7.88427(-8) \\
\ditto & 0.694 & -1.66061(0) & 8.73637(-1) & -1.52021(-1) & 1.38576(-2) & -7.05050(-4) & 1.89861(-5) & -2.11219(-7) \\ 
\ditto & 1.074 & 2.24466(0) & -9.18802(-1) & 1.76499(-1) & -1.69900(-2) & 8.96474(-4) & -2.47898(-5) & 2.81438(-7) \\
\ditto & 1.791 & -1.56419(0) & 7.86300(-1) & -1.45517(-1) & 1.38959(-2) & -7.32919(-4) & 2.03092(-5) & -2.31248(-7) \\
\ditto & 5.315 & 2.50719(-2) & -4.96470(-3) & 3.31831(-4) & -7.42211(-6) & & & \\
$13.5 < \kappa < 18.0$ & 0.452 & 1.30350(0) & -4.49713(-1) & 6.88978(-2) & -5.82877(-3) & 2.81686(-4) & -7.30064(-6) & 7.88427(-8) \\
\ditto & 0.694 & -1.66061(0) & 8.73637(-1) & -1.52021(-1) & 1.38576(-2) & -7.05050(-4) & 1.89861(-5) & -2.11219(-7) \\ 
\ditto & 1.074 & 2.24466(0) & -9.18802(-1) & 1.76499(-1) & -1.69900(-2) & 8.96474(-4) & -2.47898(-5) & 2.81438(-7) \\
\ditto & 1.791 & -1.56419(0) & 7.86300(-1) & -1.45517(-1) & 1.38959(-2) & -7.32919(-4) & 2.03092(-5) & -2.31248(-7) \\
$18.0 \leq \kappa$ & 0.411 & 4.7555(-2) & -4.1569(-3) & 1.5348(-4) & -3.0868(-6) & 3.4844(-8) & -2.0732(-10) & 5.0592(-13) \\
\ditto & 0.755 & 7.6317(-1) & -3.9928(-2) & 1.2674(-3) & -2.3391(-5) & 2.4994(-7) & -1.4319(-9) & 3.998(-12) \\
\ditto & 1.029 & -2.0600(-1) & 6.9082(-2) & -2.2280(-3) & 4.1540(-5) & -4.4691(-7) & 2.5731(-9) & -6.1319(-12) \\
\ditto & 1.551 & 3.9528(-1) & -2.4997(-2) & 8.0719(-4) & -1.5062(-5) & 1.62134(-7) & -9.3385(-10) & 2.2261(-12)
\enddata
\tablecomments{Here $c_j(\kappa) = \sum_{n}{d_{n}\kappa^{n}}$. The entries of the form a(b) are shorthand for $a\times10^{b}$.}
\end{deluxetable}

\bibliography{kappa}

\end{document}